\documentstyle[]{mn}

\newif\ifAMStwofonts

\ifoldfss
  \ifCUPmtlplainloaded \else
    \NewTextAlphabet{textbfit} {cmbxti10} {}
    \NewTextAlphabet{textbfss} {cmssbx10} {}
    \NewMathAlphabet{mathbfit} {cmbxti10} {} 
    \NewMathAlphabet{mathbfss} {cmssbx10} {} 
  \fi
  \ifAMStwofonts
    \ifCUPmtlplainloaded \else
      \NewSymbolFont{upmath} {eurm10}
      \NewSymbolFont{AMSa} {msam10}
      \NewMathSymbol{\upi}     {0}{upmath}{19}
      \NewMathSymbol{\umu}     {0}{upmath}{16}
      \NewMathSymbol{\upartial}{0}{upmath}{40}
      \NewMathSymbol{\leqslant}{3}{AMSa}{36}
      \NewMathSymbol{\geqslant}{3}{AMSa}{3E}

      \let\leq=\leqslant \let\le=\leqslant
       \let\ge=\geqslant
    \fi
  \fi
\fi 

\ifnfssone
  \newmathalphabet{\mathit}
  \addtoversion{normal}{\mathit}{cmr}{m}{it}
  \addtoversion{bold}{\mathit}{cmr}{bx}{it}
  \newmathalphabet{\mathbfit} 
  \addtoversion{normal}{\mathbfit}{cmr}{bx}{it}
  \addtoversion{bold}{\mathbfit}{cmr}{bx}{it}
  \newmathalphabet{\mathbfss} 
  \addtoversion{normal}{\mathbfss}{cmss}{bx}{n}
  \addtoversion{bold}{\mathbfss}{cmss}{bx}{n}
  \ifAMStwofonts
    \ifCUPmtlplainloaded \else
      \UseAMStwoboldmath
      \makeatletter
      \new@mathgroup\upmath@group
      \define@mathgroup\mv@normal\upmath@group{eur}{m}{n}
      \define@mathgroup\mv@bold\upmath@group{eur}{b}{n}
      \edef\UPM{\hexnumber\upmath@group}
      \new@mathgroup\amsa@group
      \define@mathgroup\mv@normal\amsa@group{msa}{m}{n}
      \define@mathgroup\mv@bold\amsa@group{msa}{m}{n}
      \edef\AMSa{\hexnumber\amsa@group}
      \makeatother
      \mathchardef\upi="0\UPM19
      \mathchardef\umu="0\UPM16
      \mathchardef\upartial="0\UPM40
      \mathchardef\leqslant="3\AMSa36
      \mathchardef\geqslant="3\AMSa3E

      \let\leq=\leqslant \let\le=\leqslant
       \let\ge=\geqslant
    \fi
  \fi
\fi 

\ifnfsstwo
  \DeclareMathAlphabet{\mathbfit}{OT1}{cmr}{bx}{it}
  \SetMathAlphabet\mathbfit{bold}{OT1}{cmr}{bx}{it}
  \DeclareMathAlphabet{\mathbfss}{OT1}{cmss}{bx}{n}
  \SetMathAlphabet\mathbfss{bold}{OT1}{cmss}{bx}{n}
  \ifAMStwofonts
    \ifCUPmtlplainloaded \else
      \DeclareSymbolFont{UPM}{U}{eur}{m}{n}
      \SetSymbolFont{UPM}{bold}{U}{eur}{b}{n}
      \DeclareSymbolFont{AMSa}{U}{msa}{m}{n}
      \DeclareMathSymbol{\upi}{0}{UPM}{"19}
      \DeclareMathSymbol{\umu}{0}{UPM}{"16}
      \DeclareMathSymbol{\upartial}{0}{UPM}{"40}
      \DeclareMathSymbol{\leqslant}{3}{AMSa}{"36}
      \DeclareMathSymbol{\geqslant}{3}{AMSa}{"3E}

      \let\leq=\leqslant \let\le=\leqslant
       \let\ge=\geqslant
    \fi
  \fi
\fi 

\ifCUPmtlplainloaded \else
  \ifAMStwofonts \else 
    \def\upi{\pi}
    \def\umu{\mu}
    \def\upartial{\partial}
  \fi
\fi

\input psfig.sty

\def\Abs#1{{\left\vert{#1}\right\vert}}

\def\R#1{{\mathrm{#1}}}		
\def\Eq#1{{Eq.~(\ref{e:#1})}}	
\def\Eqs#1#2{{Eqs.~(\ref{e:#1})-(\ref{e:#2})}}
\def\Fip#1{{~\ref{f:#1}}}       
\def\Fig#1{{Fig.~\ref{f:#1}}}   
\def\Figs#1{{Figs.~\ref{f:#1}}} 

\def\M#1{{\mathbfss{#1}}}	
\def\T#1{{#1^{\bot}}}		
\def\d#1{{\R{d}{#1}}}		
\def\mdot{\!\cdot\!}		

\def\StackUnder#1#2{\mathrel{\mathop{#1}\limits_{#2}}}

\def\SNR{{S\!N\!R}}
\def\eps{{\varepsilon}}
\def\epsmin{{\eps_\R{min}}}
\def\etacrc{{\eta_\R{c}}}

\def\Hmax{{h_\R{max}}}
\def\paral{{/\!/}}

\def\iint{\int\!\!\!\!\!\int}	

\def\Trash#1{{}}

\title[Distribution functions from observed galactic disks]{Non parametric
 reconstruction of distribution functions\\ from observed galactic disks}

\author[C.~Pichon \& E.~Thi\'{e}baut]
       {C.~Pichon${}^{1,2,3}$\\
	${}^1$ CITA, 60 St. George Street, Toronto, Ontario M5S 1A7, Canada. \\
	${}^2$ Astronomisches Institut, Universitaet Basel, Venusstrasse 7,
	CH-4102 Binningen Switzerland\\
	${}^3$ Institut d'Astrophysique de Paris, 98 bis boulevard
       d'Arago, 75014 Paris, France.
	\newauthor
	E.~Thi\'{e}baut\\
        Centre de Recherches Astronomiques de Lyon, 9 avenue Charles
	Andr\'{e}, F-69561 Saint Genis Laval Cedex}

\date{\today}

\pagerange{\pageref{firstpage}--\pageref{lastpage}}
\pubyear{1997}
\begin{document}

\maketitle

\label{firstpage}


\begin{abstract}
A  general inversion technique   for the  recovery of  the  underlying
distribution  function for observed  galactic  disks is presented  and
illustrated.  Under the assumption that  these disks are axi-symmetric
and   thin,   the proposed  method     yields the unique  distribution
compatible with all the observables  available.  The derivation may be
carried   out   from   the measurement    of   the azimuthal  velocity
distribution arising from positioning the slit of a spectrograph along
the major axis of the galaxy.  More  generally, it may account for the
simultaneous measurements of  velocity  distributions corresponding to
slits  presenting  arbitrary orientations  with  respect  to the major
axis.  The approach  is non-parametric,  i.e.\ it  does not rely  on a
particular algebraic model  for  the distribution  function.   Special
care is taken to  account for the  fraction of counter-rotating  stars
which strongly affects the stability of the disk.

An optimisation algorithm  is devised   --  generalising the work   of
Skilling  \&   Bryan (1984) --  to    carry this truly two-dimensional
ill-conditioned   inversion  efficiently.    The  performances of  the
overall  inversion technique   with respect to  the    noise level and
truncation in the  data   set  is investigated  with  simulated  data.
Reliable results are obtained up to a  mean signal to noise ratio of~5
and when measurements are available up to $4  R_{e}$.  A discussion of
the residual  biases    involved  in  non  parametric   inversions  is
presented.  Prospects of  application  to observed galaxies and  other
inversion problems are discussed.
\end{abstract}


\section{Introduction}

In  years to come, accurate  kinematical measurement of nearby disk galaxies
will be achievable with  high  resolution spectroscopy.  Measurement  of the
observed  line  profiles will yield  relevant  data to  probe the underlying
gravitational nature of the interaction holding the galaxy together.  Indeed
the  assumption that the   system is stationary  relies  on the existence of
invariants which   put  severe  constraints    on   the  possible   velocity
distributions.  This is formally expressed by the existence of an underlying
distribution  function which    specifies the   dynamics  completely.    The
determination of realistic distribution functions which account for observed
line profiles is therefore required in order  to understand of the structure
and dynamics of spiral galaxies.

Inversion methods have been implemented for  spheroids (globular clusters or
elliptical    galaxies)  by   Merrifield  \shortcite{Merrifield91}, Dejonghe
\shortcite{Dejonghe93},   Merritt \shortcite{DM96} \shortcite{DM97}, Merritt
\& Tremblay \shortcite{MerrittTremblay94} \shortcite{DM87}, Emsellem, Monnet
\&  Bacon  \shortcite{EmsellemEtAl94},  Dehnen \shortcite{Dehnen95}, Kuijken
\shortcite{Kuijken95},  and Qian  \shortcite{Qian95}.  Indeed for spheroids,
the surface  density  alone  yields  access to   the  even  component  of  a
2-integral distribution function which may account for the internal dynamics
(while the  odd component can be   recovered from the  mean azimuthal flow).
However the  corresponding  recovered distribution might   not be consistent
with higher Jeans moments, since  the equilibria may involve three (possibly
approximate) integrals.  The  inversion problem corresponding to a flattened
spheroid which  is assumed to have 2  or 3 (Stakel-based) integrals has been
addressed  recently  by  Dejonghe   et al.\  \shortcite{DejongheEtAl96}  and
illustrated on NGC~4697.  Non parametric  approaches have in particular been
used with success  by Merritt \& Gebhardt  \shortcite{MerrittGebhardt94} and
Gebhardt \& al.  \shortcite{Faber}   to solve the dynamical inverse  problem
for the density in spherical geometry.  If the spheroid is seem exactly edge
on Merritt \shortcite{DM96} has devised a method which allow them to recover
simultaneously the underlying potential.

Here the inversion problem for  thin  and  round  disks  is  addressed where
symmetry ensures integrability.  In this context, the  inversion  problem is
truly two-dimensional and requires special attention  for  the  treatment of
quasi-radial orbits in the inner part of the galaxy.

By  Jeans' theorem  the   steady-state mass-weighted   distribution function
describing a flat galaxy must be of the form  $f = f(\varepsilon, h)$, where
the specific energy, $\varepsilon$, and  the specific angular momentum, $h$,
are given by
\begin{equation}
  \varepsilon = {\textstyle{1\over2}} (v_R^2 + v_\phi^2) - \psi \, , \quad
  \mbox{and} \quad h = R\, v_\phi \, .
\end{equation}
\noindent Here $v_R$ and $v_\phi$ are the star radial and angular velocities
respectively of stars confined to a plane and $\psi(R)$ is the gravitational
potential  of     the   disk.    The   azimuthal    velocity   distribution,
$F_\phi(R,v_\phi)$, follows from this distribution according to:
\begin{equation}
	F_\phi(R,v_\phi)= \int f(\eps,h)\,\d{v_R} \, ,
	\label{e:Fphi-1} 
\end{equation}
\noindent where  the  integral  is  over  the  region ${\textstyle{1\over2}}
(v_R^2 + v_\phi^2)  <  \psi$  corresponding  to  bound  orbits.   Pichon and
Lynden-Bell \shortcite{PchLBd96} demonstrated that, in the  case  of  a thin
round galactic disk, the distribution can be analytically inverted  to yield
a unique  $f(\eps,h)$  provided  the  potential  $\psi(R)$  is  known.   The
velocity  distribution  $F_\phi(R,v_\phi)$  can  be  estimated  --  within a
multiplicative constant -- from line of sight velocity  distribution (LOSVD)
data obtained by long slit spectroscopy when the slit  is  aligned  with the
major axis of the galactic disk  projected  onto  the  sky.   Similarly, the
rotation curve observed in HI gives in principle  access  to  the underlying
potential.    More   generally,   simultaneous   measurements   of  velocity
distributions are derived with slits presenting arbitrary  orientations with
respect to the major axis as discussed in Appendix~\ref{section:alpha}.


The inversion  of  \Eq{Fphi-1}  is  known  to  be  ill-conditioned:  a small
departure in the  measured  data  (e.g.\  due  to  noise)  may  produce very
different solutions since these are dominated by artifacts  corresponding to
the amplification of noise.  Some kind of balance  must  therefore  be found
between the constraints imposed on the solution in order to deal  with these
artifacts on the one hand, and the degree  of  fluctuations  consistent with
the assumed information contents of the signal on the other hand  (i.e.\ the
worse the data  quality  is,  the  lower  the  informative  contents  of the
solution will be and the more constraint the restored  distribution  will be
so as to avoid an over-interpretation of the data ).  Finding such a balance
is  called  the  ``regularisation''   of   the   inversion   problem  (e.g.\
\cite{Wahba79}) and methods implementing  adaptive  level  of regularisation
are described as ``non parametric''.
 
Under the assumption  that  these  disks  are  axi-symmetric  and  thin, the
proposed non parametric methods described in this paper yield in principle a
unique  distribution:  the  smoothest  solution  consistent  with   all  the
available  observables,  the  knowledge  of  the  level  of  noise  in  each
measurement and some  objective  physical  constraints  that  a satisfactory
distribution should fulfill.

Section~\ref{section:nonparam}  presents all relevant theoretical aspects of
regularisation     and   non    parametric  inversion  for   galactic  disks
distributions.    Section~\ref{section:algo}  present the various algorithms
and the corresponding numerical techniques which  we implemented in steps to
carry efficiently this   two-dimensional  minimisation.  It corresponds   in
essence     to  an     extension    of   the    work     of  Skilling     \&
Bryan\shortcite{SkillingBryan84} for maximum  entropy  to  other  penalising
functions  which are  more relevant  in   this context.   All techniques are
implemented  in section~4 on   simulated data arising  when the  slit of the
spectrograph is  aligned  with the  long  axis of  the   projected disk.   A
discussion follows.

\section{Non-parametric inversion for flat \& round disks}
\label{section:nonparam}

The non parametric inversion problem involves finding the  best  solution to
\Eq{Fphi-1} for the distribution function when  only  discretised  and noisy
measurements of $ F_\phi(R,v_\phi)$ are available.

A  distinction  between  {\em  parametric\/}   and   {\em  non-parametric\/}
descriptions may  seem  artificial:  it  is  only  a  function  of  how many
parameters are needed to describe the model with respect  to  the  number of
independent measurements.  In a parametric model there are a small number of
parameters compared to the number of data samples.  This makes the inversion
for the parametric model somewhat regularised, i.e.\ well-conditioned.  Yet,
once the model has been chosen, there is no  way  to  control  the  level of
regularisation and the inversion will always produce a solution, whether the
parametric model and its implicit level of regularisation is correct or not.
In a non-parametric model, as a result of the discretisation, there  is also
a finite number of parameters but it is comparable and  usually  larger than
the number of  data  samples.   In  this  case,  the  amount  of information
extracted from the data is  controlled  explicitely  by  the regularisation.
Here the  latter  non  parametric  method  is  therefore  prefered  since no
particular unknown physical model for disks distributions is to be favoured.

\subsection{The discretised kinematic integral equation}

Since $\eps$ is an even function of $v_R$  and  since  the  relation between
$v_R$ and $\eps$ is one-to-one on the  interval  $v_R\in[0,\infty)$  and for
given $R$ and $v_\phi$, Eq.~(\ref{e:Fphi-1}) can be rewritten explicitly as:
\begin{equation}
  F_\phi(R,v_\phi)   =   \sqrt{2}\!\!\!\!\int\limits_{-Y(R,v_\phi)}^{0}\!\!
  	\frac{f(\eps,R\,v_\phi)}{\sqrt{\eps+Y(R,v_\phi)}}\,\d{\eps}
  \, ,	\label{e:Fphi-2} 
\end{equation}
\noindent where the effective potential is given by: 
\begin{equation}
  Y(R,v_\phi) = \psi(R)-\frac{1}{2}v^2_\phi \, .  \label{e:Y} 
\end{equation}
For a given angular momentum $h$ the minimum specific energy is:
\begin{equation}
  \epsmin(h) = \StackUnder{\R{min}}{R\in[0,\infty)}
	\Bigl\{\frac{h^2}{2R^2}-\psi(R)\Bigr\} \, .  \label{e:epsmin} 
\end{equation}
\noindent From  \Eq{Fphi-2},  the  generic  ill-conditioning  of \Eq{Fphi-1}
appears clearly since the integral relation relating the  azimuthal velocity
distribution and the underlying distribution is an Abel  transform  (i.e.\ a
half derivative).

Given the error level in the measurements  and  the  finite  number  of data
points $N_\R{data}$, $f(\eps,h)$ is derived by fitting  the  data  with some
model.  Since the number of physically relevant distributions $f(\eps,h)$ is
very large, a  small  number  of  parameters  cannot  describe  the solution
without further assumptions (i.e.\ other than that  the  disk  is  round and
thin).  A general approach must  therefore  be  adopted;  for  instance, the
solution can be described  by  its  projection  onto  a  basis  of functions
$\{e_k(\eps,h); k=1,\ldots,N\}$:
\begin{equation}
	f(\eps,h) = \sum_{k=1}^N f_k \; e_k(\eps,h) \;. 
	\label{e:f-model-1}
\end{equation}
The parameters to fit are the weights $f_k$. In order to fit a  wide variety
of functions, the basis must be very large; consequently the  description of
$f(\eps,h)$ is no longer parametric  but  rather  non-parametric.

In order to account for the fact that the equilibrium should not incorporate
unbound  stars  it  is  best  to  define  the  functions   $e_k(\eps,h)$  of
\Eq{f-model-1} so  that  they  are  identically  zero  outside  the interval
$(\eps,h)\in[\epsmin(h),0]\times\R{I\!R}$.  It is convenient to rectify this
interval  while  replacing  the   integration   over   specific   energy  in
Eq.~(\ref{e:Fphi-2}) by an integration with respect to:
\begin{equation}
	\eta = 1 - \frac{\eps}{\epsmin(h)} \; , 
\end{equation}
and to use a new basis of functions: 
\begin{displaymath}
	\hat{e}_k(\eta, h) \equiv
	e_k\bigl((1\!-\!\eta)\,\epsmin\!(h),\,h\bigr)  \, ,
\end{displaymath}
which are zero outside the interval $(\eta,h)\in[0,1]\times\R{I\!R}$.   Here
$\eta$ is some measure of the eccentricity of  the  orbit.   Using  this new
basis functions, the distribution function becomes:
\begin{equation}
	\hat{f}(\eta,h) \equiv f\bigl((1\!-\!\eta)\,\epsmin\!(h),\,h\bigr)
	= \sum_{k=1}^N f_k \; \hat{e}_k(\eta,h) \,.
	\label{e:f-model-2}
\end{equation}
Another  important  advantage  of  this   reparametrisation   is   that  the
distributions $\hat{f}(\eta,h)$ can  be  assumed  to  be  smoother functions
along $\eta$ and $h$ since these distributions correspond to  the equilibria
of relaxed and cool system which have gone  through  some  level  of violent
relaxation in their formation processes and  where  most  orbits  are almost
circular. Note nonetheless that this assumption is  somewhat  subjective and
introduces some level of bias corresponding to what is  considered  to  be a
good    distribution     function     as     will     be     discussed    in
section~\ref{s:discussion}.  Clearly the  assumption  that  the distribution
function should be smooth (i.e.  without strong gradients)  in  the variable
$\eta$ yields different constraints on the sought solution than  assuming it
should be smooth in the variable $\varepsilon$.

Real data correspond to discrete measurements $R_i$ and ${v_\phi}_j$  of $R$
and  $v_\phi$  respectively.   Following  the  non-parametric  expansion  in
Eq.~(\ref{e:f-model-2}), Eq.~(\ref{e:Fphi-2}) now becomes:
\begin{equation}
	F_{i,j} \equiv F_\phi(R_i,{v_\phi}_j) =  \sum_{k=1}^N  a_{i,j,k} \;
		f_k \, , \label{e:Fphi-model} 
\end{equation}
with
\begin{equation}
 	a_{i,j,k} = \sqrt{-2\epsmin_{i,j}} \int^1_{\etacrc_{i,j}}
	\frac{\hat{e}_k(\eta, R_i\,{v_\phi}_j)}
	{\sqrt{\eta-\etacrc_{i,j}}}\,\d{\eta}\,.
	\label{e:Fphi-model-coef} 
\end{equation}
where
\begin{equation}
	\epsmin_{i,j}\equiv\epsmin(R_i\,{v_\phi}_j) \, , \quad 
	\etacrc_{i,j}\equiv 1+Y(R_i,{v_\phi}_j)/\epsmin_{i,j}\,.
\end{equation} 
The implementation of this linear  transformation  for  linear  B-splines is
given in Appendix~A.  Since  the  relations  between  $F_\phi(R,v_\phi)$ and
$f(\eta,h)$ or $\hat{f}(\eta,h)$ are linear, Eq.~(\ref{e:Fphi-model}) -- the
discretised form of the integral equation~(\ref{e:Fphi-1}) -- can be written
in a matrix form by grouping index $i$ with index $j$:
\begin{equation}
	\M{F} = \M{a} \mdot \M{f} \,.
	\label{e:Fphi-model-2}
\end{equation}
The problem of solving \Eq{Fphi-1} becomes a linear inversion problem.

\subsection{Maximum Penalised Likelihood}

In order to model a wide range of distributions $\hat{f}(\eta,h)$  with good
accuracy,   the   basis   $\{\hat{e}_k(\eta,h);   k=1,\dots,N\}$   must   be
sufficiently general (otherwise the solutions will be biased  by  the choice
of the basis just as a parametric approach is biased by  the  choice  of the
model). The inversion should therefore be regularised and performed so as to
avoid  physically  irrelevant  solutions.   Indeed,  being  a  distribution,
$\hat{f}(\eta,h)$ must for instance be positive  and  normalised.   Finally,
the inversion should provide some level of flexibility  to  account  for the
fact that the sought distribution might have a critical  behaviour  for some
fraction of phase space such as that corresponding  to  radial  orbits.   It
should also cope with incomplete data sets and should  yield  some  means of
extrapolation.

In order to address these specificities let us  explore  techniques  able to
perform a reliable practical inversion of this  ill-conditioned  problem and
put the method brought forward in this  paper  into  context.   The Bayesian
description provides a  suitable  framework  to  discuss  how  the practical
inversion of \Eq{Fphi-model-2} should be performed.

\subsubsection{Bayesian approach}

When dealing with real data, noise must be  accounted  for:  instead  of the
exact solution of \Eq{Fphi-model},  it  is  more  robust  to  seek  the {\em
best\/}  solution  compatible  with  the  data  and,   possibly,  additional
constraints.  A criteria allowing to select such a solution  is  provided by
probability analysis.  Indeed, given the measured data  $\tilde{\M{F}}$, one
would like to recover the  most  probable  underlying  distribution $\M{f}$.
This is achieved by maximising the probability of  the  distribution $\M{f}$
given  the  data  $\tilde{\M{F}}$,  $\Pr(\M{f}\,\vert\,\tilde{\M{F}})$, with
respect     to     $\M{f}$.       According      to      Bayes'     theorem,
$\Pr(\M{f}\,\vert\,\tilde{\M{F}})$, can be rewritten as:
\begin{equation}
	\Pr(\M{f}\,\vert\,\tilde{\M{F}})=
	\frac{\Pr(\tilde{\M{F}}\,\vert\,\M{f})\Pr(\M{f})}
	{\Pr(\tilde{\M{F}})} \, , 
\end{equation}
\noindent where $\Pr(\tilde{\M{F}}\,\vert\,\M{f})$ is the probability of the
data $\tilde{\M{F}}$ given that it  should  obey  the  distribution $\M{f}$,
while  $\Pr({\tilde{\M{F}}})$   and   $\Pr(\M{f})$   are   respectively  the
probability  of  the  data  $\tilde{\M{F}}$  and  the  probability   of  the
distribution  $\M{f}$.   Since  $\Pr(\tilde{\M{F}})$  does  not   depend  on
$\M{f}$,  maximising  $\Pr(\M{f}\,\vert\,\tilde{\M{F}})$  with   respect  to
$\M{f}$ is equivalent to minimising:
\begin{equation}
	Q(\M{f}) = L(\M{f}) + \mu R(\M{f}) \, ,	\label{e:Q}
\end{equation}
with
\begin{eqnarray}
	L(\M{f}) &=& - \alpha \mathrm{Log}[\Pr(\tilde{\M{F}}\,\vert\,\M{f})]
		+ c \, , \label{e:L} \\
	\mu R(\M{f}) &=& - \alpha \mathrm{Log}[\Pr(\M{f})]
		+ c' \, , \label{e:R} 
\end{eqnarray}
with $\alpha>0$ and where $c$ and $c'$ are constants which  account  for any
contribution which does not depend on $\M{f}$.   Minimising  the likelihood,
$L(\M{f})$, enforces consistency of the model with the data while minimising
$R(\M{f})$ tends to give the ``most  probable  solution''  when  no  data is
available as discussed in section~\ref{sect:reg}.

\subsubsection{Maximum likelihood}

Minimisation of $L(\M{f})$ alone in  \Eq{Q}  yields  the  maximum likelihood
solution.            The            exact            expression           of
$-\mathrm{Log}[\Pr(\tilde{\M{F}}\,\vert\,\M{f})]$ can usually be derived and
depends on the noise statistics.  For instance, assuming that  the  noise in
the measured data follows a normal law,  maximising  the  likelihood  of the
data is obtained by minimising the $\chi^2$ of the data:
\begin{displaymath}
	-\mathrm{Log}[\Pr(\tilde{\M{F}}\,\vert\,\M{f})]
	= \frac{1}{2}\chi^2 + c'' \quad \mbox{with} \quad
	\chi^2 \equiv \sum_{i,j}\frac{(F_{i,j}-\tilde{F}_{i,j})^2}
		{\mathrm{Var}(\tilde{F}_{i,j})} \, , 
\end{displaymath}
\noindent   where   $F_{i,j}$   is   the   model   of   $F_\phi$   given  by
Eq.~(\ref{e:Fphi-model})  and  $\tilde{F}_{i,j}$  denotes  the  measures  of
$F_\phi$.  Minimisation of $\chi^2$ is known  as  {\em  Chi-square fitting}.
Throughout this paper and for the sake of clarity, Gaussian
noise is  assumed while defining the likelihood term by:
\begin{equation}
	L(\M{f}) = \chi^2(\M{f}) = \sum_{i,j}
	\frac{(F_{i,j}-\tilde{F}_{i,j})^2}{\mathrm{Var}(\tilde{F}_{i,j})}
	\,,	\label{e:L=chi2}
\end{equation}
(which incidentally corresponds to the choice $\alpha=2$ in Eq.~(\ref{e:L})).
In the limit of a large number of independent  measurements, $N_\R{data}$,
$\chi^2$ follows a normal law with an expected value and a  variance given
by:
\begin{equation}
	\mathrm{Expect}(\chi^2) = N_\R{data} \, , \quad \R{and} \quad
	\mathrm{Var}(\chi^2) = 2\,N_\R{data} \nonumber\, . 
\end{equation}
\noindent It follows that any distribution,  $\M{f}$,  yielding  a  value of
$\chi^2$  in  the  range  $N_\R{data}\pm\sqrt{2\,N_\R{data}}$,  is perfectly
consistent with the measured data: none of those distributions  can  be said
to be better than others on the basis of the measured data alone.

For a parametric description and provided that the number  of  parameters is
small compared to $N_\R{data}$, the region around the minimum of $\chi^2$ is
usually very narrow.  In this case, Chi-square fitting  may  be sufficiently
robust to produce a reliable solution (though this conclusion depends on the
noise level and assumes that the parametric model is correct).

In a non-parametric approach, given the  functional  freedom  left  in the
possible distributions, it is likely that the value of the  $\chi^2$  can be
made arbitrarily small, i.e.\ much smaller than $N_\R{data}$.  Consequently,
the solution which minimises $\chi^2$ is not reliable: it is too good  to be
true!  In  other  words,  solely  minimising  $\chi^2$  in  a non-parametric
description  leads  to  an  over-interpretation  of  the  data:  due  to the
ill-conditioned nature of the problem, many  features  in  the  solution are
likely to be artifacts  produced  by  amplification  of  noise  or numerical
rounding errors.

\subsubsection{Regularisation\label{sect:reg}}

Minimising the likelihood term forces the model to be  consistent  with some
objective  information:  the  measured  data.   Nevertheless,  this approach
provides no means of selecting a particular solution among  all  those which
are    consistent    with    the    data    (i.e.\    those     for    which
$L(\M{f})=N_\R{data}\pm\sqrt{2\,N_\R{data}}$).     Taking    into    account
$\mu{}R(\M{f})=-\alpha\mathrm{Log}[\Pr(\M{f})]+c''$  in   \Eq{Q}   yields  a
natural procedure to choose between those solutions.   At  least,  there are
some objective properties of the  distribution  $\hat{f}(\eta,h)$  which are
not enforced by Chi-square fitting (e.g.\  positivity)  and  which  could be
accounted  for  by  the  fact  that  $\Pr(\M{f})$   must   be   zero  (i.e.\
$R(\M{f})\rightarrow\infty$) for physically irrelevant solutions.

Unfortunately, e.g.\ for noisy data,  taking  into  account  those objective
constraints alone is seldom sufficient:  additional  ad-hoc  constraints are
needed to regularize the inversion  problem.   To  that  end,  $R(\M{f})$ is
generally defined as a so-called {\em penalising  function}  which increases
with the discrepancy between $\M{f}$ and those {\em subjective constraints}.

To summarise, the solution of Eq.~(\ref{e:Fphi-1})  is  found  by minimising
the   quantity   $Q(\M{f})=L(\M{f})+\mu\,R(\M{f})$   where   $L(\M{f})$  and
$R(\M{f})$ are respectively  the  likelihood  and  regularisation  terms and
where the parameter $\mu>0$ allows to tune the level of regularisation.  The
introduction  of  the  Lagrange  multiplier  $\mu$  in  \Eq{Q}  is  formally
justified by the fact that  $Q(\M{f})$ should be minimised   subject  to the
constraint that $L(\M{f})$ should be equal  to  some  value,  say $N_\R{e}$.
For instance, with $L(\M{f})=\chi^2(\M{f})$ one would choose
\begin{displaymath}
    N_\R{e} \in [N_\R{data} - \sqrt{2\,N_\R{data}},
                 N_\R{data} + \sqrt{2\,N_\R{data}}]\,.
\end{displaymath}

\subsubsection{Definitions of the penalising function}

When data consist in samples of a continuous  physical  signal, uncorrelated
noise will  contribute  to  the  roughness  of  the  data.   Moreover, noise
amplification by an ill-conditioned inversion is likely to produce  a forest
of  spikes  or  small  scale  structures  in  the  solution.   As  discussed
previously, assuming that the  ``probability''  $\Pr(\M{f})$  increases with
the smoothness of $\hat{f}(\eta,h)$, the  penalising  function  should limit
the effects of noise while not affecting (i.e.\ biasing) too much  the range
of  possible  shape  of  $\hat{f}(\eta,h)$.   To  that  end,  the penalising
function $R(\M{f})$ should be defined so  as  to  measure  the  roughness of
$\M{f}$.

Many different penalising functions can be defined to measure  the roughness
of $\hat{f}(\eta,h)$; for instance minimising \cite{Wahba90}:
\begin{equation}
	R(\M{f}) = \iint \biggl[\!
	\M{\nabla}^n  \hat{f} \mdot \M{\nabla}^n \hat{f}	
\biggr] \d{\eta}\,\d{h}\, ,  \quad \!\! { \rm with} \quad \!\! 
\M{\nabla} =  \biggl(\!\frac{\partial\!\hat{f}}{\partial h},
\!\!\frac{\partial\!\hat{f}}{\partial\eta}  \biggl)
	\label{e:Rquad-general}
\end{equation}


\noindent (where $n>1$) will enforce  the  smoothness  of $\hat{f}(\eta,h)$.
In the instance of a discretised  signal  for  Eq.~(\ref{e:f-model-2}), such
quadratic penalising functions can be generalised by the use  of  a positive
definite operator $\M{K}$ \cite{Titterington85}:
\begin{equation}
	R_\R{quad}(\M{f}) = \T{\M{f}} \mdot \M{K} \mdot \M{f} \,,
	\label{e:Rquad}
\end{equation}
where $\T{\M{f}}$ stands for the transpose of $\M{f}$.


Strict application  of  the Bayesian analysis   implies that  the penalising
function  $R(\M{f})$   is  $-\mathrm{Log}[\Pr(\M{f})]$  (up   to an additive
constant and the   factor $\mu$) which is   the negative of the  entropy  of
$\M{f}$.  This  has led to the family  of maximum entropy methods (hereafter
MEM)  which are widely used  to solve ill-conditioned  inverse problems.  In
fact MEM  only differs from   other  regularised methods by  the  particular
definition of the penalising function which provide positivity ab initio.  A
possible definition of the negentropy is \cite{Skilling89}:
\begin{equation}
	R_\R{MEM}(\M{f}) =
		\sum_k \bigl[f_k\,\R{Log}\frac{f_k}{p_k} - f_  k + p_k\bigr]
	\,,\label{e:Rmem}
\end{equation}
where $\M{p}$ is the {\em a priori} solution: the entropy is  maximised when
$\M{f}=\M{p}$.  Although there are arguments in  favour  of  that particular
definition,     there     are      many      other      possible     options
\cite{NarayanNityananda1986} which lead to  similar  solutions.   Penalising
functions in MEM all share the property that they become infinite as $\M{f}$
reaches zero, thus enforcing positivity.  In order  to  further  enforce the
smoothness of the solution, Horne \shortcite{Horne85} has suggested  the use
of a floating  prior,  defining  $\M{p}$  to  be  $\M{f}$  smoothed  by some
operator $\M{S}$:
\begin{equation}
	\M{p}=\M{S}\mdot\M{f}\,.
	\label{e:floating-prior}
\end{equation}
For instance, along {\em each dimension} of $\hat{f}(\eta,h)$, the
following mono-dimensional smoothing operator is applied:
\begin{displaymath}
	p_i= \left\{\begin{array}{ll}
	(1-\gamma) f_i + \gamma f_{i+1}\,, & \mbox{if $i=1$}\,,\\
	\gamma f_{i-1} + (1-2\gamma) f_i + \gamma f_{i+1}\,, &
		\mbox{if $1<i<n$}\,,\\
	(1-\gamma) f_i + \gamma f_{i-1}\,, & \mbox{if $i=n$}\,,\\
	\end{array}\right.
\end{displaymath}
with $0\le\gamma\le1/2$ (here: $\gamma=1/4$); here  $i=1,...,n$  stands for
the index along the dimension considered.  This  operator  conserves energy,
i.e.\ $\sum{}p=\sum{}f$.

The penalising functions $R_\R{quad}(\M{f})$  or  $R_\R{MEM}(\M{f})$  with a
floating prior $\M{p}=\M{S}\mdot\M{f}$ is implemented in the  simulations to
enforce the smoothness of the solution.

\subsubsection{Adjusting the weight of the regularisation}

Thompson and Craig \shortcite{ThompsonCraig92} compared many  different {\em
objective methods} to fix the actual value  of  $\mu$.   Generally speaking,
these methods consists in minimising $Q(\M{f})$ given by  \Eq{Q}  subject to
the constraints $L(\M{f})=N_\R{e}$  where  $N_\R{e}$  is  equivalent  to the
number of degrees of freedom of the model.  Among those methods, two  can be
applied to non-quadratic penalising functions (such as the negentropy).

The most simple approach is to minimise $Q(\M{f})$ subject to the constraint
that  $L(\M{f})=\R{Expect}[L(\M{f})]=N_\R{data}$.   This   yields   an  {\em
over-regularised} solution \cite{Gull89} since it is equivalent  to assuming
that regularisation controls no degrees of freedom.

A second method is due to Gull \shortcite{Gull89} who demonstrated  that the
Lagrange parameter should  be  tuned  so  that  $Q(\M{f})=N_\R{data}$, i.e.\
$N_\R{e}=N_\R{data}-\mu{}R(\M{f})$.  In other words, the sum  of  the number
of degrees of freedom controlled by the data and by the entropy is  equal to
the number of measurements.  This method is very simple to implement but can
lead  to  {\em  under-regularised}  solutions \cite{Gull89,ThompsonCraig92}.
Indeed if the subjective constraints pull  $\M{f}$  too  far  from  the true
solution then $R(\M{f})$ takes a high value as soon as any structure appears
in $\hat{f}(\eta,h)$.  As a result, in order to  meet $Q(\M{f})=N_\R{data}$,
the value of $\mu$ is found  to  be  very  small  by  this  procedure.   For
instance, this occurs in MEM methods when choosing a  uniform  prior $\M{p}$
since a  uniform  distribution  is  very  far  from  the  true distribution.
Nevertheless,this kind of problem was not encountered with a  floating prior
\cite{Horne85}.  In the algorithm described below this latter  method (i.e.\
Gull plus Horne methods) is  implemented  to  obtain  a  sensible  value for
$\mu$.

Another potentially attractive  way  to  find  the  value  of  $\mu$  is the
cross-validation method \cite{Wahba79} since it relies  solely  on  the data.
Let $\dot{F}_{i,j}$  be the value  at $(i,j)$  of  the model which  fits the
subset of data derived while excluding measurement  $(i,j)$ (in other words,
$\dot{F}_{i,j}$ predicts the    value of  the  assumed   missing data  point
$\tilde{F}_{i,j}$); since the fit is achieved  by minimising $Q(\M{f})$, the
total prediction error, given by:
\begin{displaymath}
	T\!P\!E = \sum_{i,j}\frac{[\dot{F}_{i,j}-\tilde{F}_{i,j}]^2}
	{\R{Var}(\tilde{F}_{i,j})}
\end{displaymath}
will depend on  the sought value  of $\mu$.  The  so-called cross-validation
method chooses the value of $\mu$  that minimise $T\!P\!E$.  When the number
of   data points  is  large this  method  becomes  too {\sc  cpu} intensive.
Nonetheless   Wahba     \shortcite{Wahba90}  and    also        Titterington
\shortcite{Titterington85}  provide  efficient means  of choosing $\mu$ when
the model is  linear which involve  constructing  the so called generalised
cross validation  estimator for the $T\!P\!E$.

\begin{figure*}
\centerline{\psfig{file=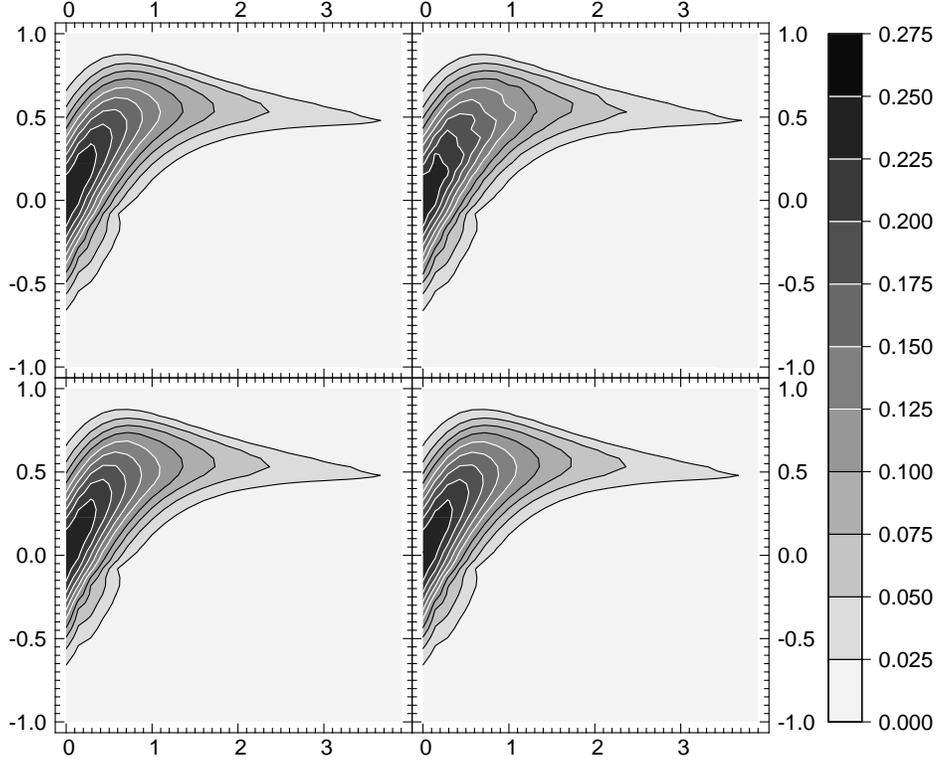,width=1.5\columnwidth}}
\caption{Fit  of  $F_\phi$  (as  described  in  section~\ref{sect:sim}) with
various penalizing functions.  From top-left to  bottom-right:  (1) original
$F_\phi$ and $F_\phi$ restored by (2) MEM with uniform prior,  (3)  MEM with
floating smooth prior and (4) quadratic  regularisation,  i.e.\ $R_\R{quad}$
with $n=1$ as defined in Eq.~\protect\ref{e:Rquad-general}.  As expected, no
significant difference is to be found in  the  fits,  though  in  panel (2),
$F_\phi$ is slightly rougher.  The SNR is~50.\label{f:fig1}}
\end{figure*}

\begin{figure*}
\centerline{\psfig{file=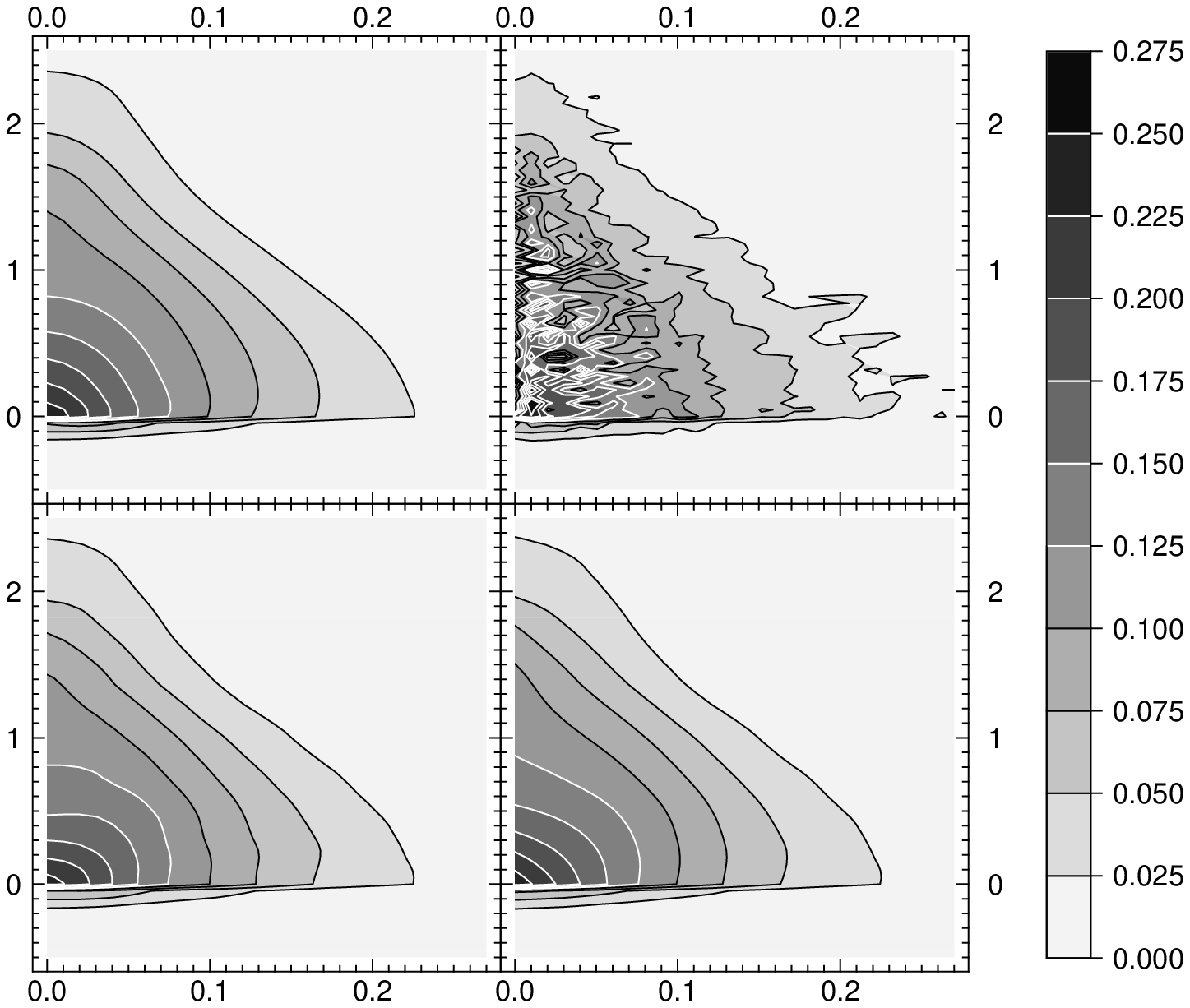,width=1.5\columnwidth}}
\caption{Restoration of $\hat{f}(\eta,h)$  from  \protect{\Fig{fig1}} (from
top-left to bottom-right): (1) true distribution and distributions restored
by (2) MEM with uniform prior, (3) MEM with smooth floating  prior  and (4)
quadratic regularisation,  i.e.\  $R_\R{quad}$  with  $n=1$  as  defined in
Eq.~\protect\ref{e:Rquad-general}.  Note  that  MEM  with  a  uniform prior
yields a rather unsmooth solution, which is expected  since  no  penalty is
imposed by this method for lack of smoothness.\label{f:fig2}}
\end{figure*}

\section{Numerical optimisation}\label{section:algo}

In the previous section it was  shown that the inversion problem  reduces to
the      minimisation       of       a       multi-dimensional      function
$Q(\M{f})=L(\M{f})+\mu{}R(\M{f})$  with  respect  to  a   great   number  of
parameters (from a few $10^4$ to $10^6$) and subject to the constraints that
(i) the likelihood term keeps some  target  value:  $L(\M{f})=N_\R{e}$, (ii)
all parameters remain positive and that (iii) special  care  is  taken along
some physical boundaries.  Unfortunately there exists  no  general black-box
algorithm able to perform this kind of optimisation.

Let  us  therefore   investigate in   turns three techniques   to  carry the
minimisation  of  increasing efficiency   and complexity:   direct  methods,
iterative  minimisation along a single  direction (accounting for positivity
at  fixed regularisation)  and    iterative  minimisation with  a   floating
regularisation weight.

\subsection{Linear solution}\label{section:linear}

Using quadratic regularisation, the problem is solved by minimizing:
\begin{equation}
	Q_\R{quad}(\M{f}) =
		\T{(\tilde{\M{F}} - \M{a}\mdot \M{f})} \mdot \M{W}
		\mdot (\tilde{\M{F}} - \M{a} \mdot \M{f})
		+ \mu \, \T{\M{f}} \mdot \M{K} \mdot \M{f} \, .
	\label{e:Q-quad}
\end{equation}
\noindent where $\M{W}$ is the inverse of the covariance matrix of the data.
The solution $\M{f}_\R{quad}$ which minimises $Q_\R{quad}$ is:
\begin{equation}
	\M{f}_\R{quad} = (\T{\M{a}} \mdot \M{W} \mdot \M{a} +
		\mu\,\M{K})^{-1} \mdot \T{\M{a}} \mdot \M{W} \mdot \M{F}\,.
		\label{e:Q-quad-solution}
\end{equation}
This solution, which is linear with respect  to  the  data,  is  clearly not
constrained to be positive.

\subsection{Non linear optimisation} 

Linear methods  only provide raw, possibly locally  negative, solutions.  At
the very  least, enforcing positivity of the  solution -- and more generally
if   the   penalised  function  is not   quadratic    -- requires non-linear
minimisation.  In that case, the minimisation  of $Q(\M{f})$ must be carried
out by successive approximations.

At the $n^{\mathrm{th}}$ step, such iterative  minimisation  methods usually
proceed by varying the current parameters  $\M{f}^{(n)}$  along  a direction
$\delta\M{f}^{(n)}$  so  as  to  minimise  $Q$;  the  new  estimate  of  the
parameters reads:
\begin{equation}
	\M{f}^{(n+1)}=\M{f}^{(n)}+\lambda^{(n)} \, \delta\M{f}^{(n)} \,,
\end{equation}
where  the optimum step size $\lambda^{(n)}$ is the scalar:
\begin{equation}
	\lambda^{(n)} = \arg \{ \min_{\lambda}[
	Q(\M{f}^{(n)}+\lambda \, \delta\M{f}^{(n)})] \} \,.
	\label{e:monocp}
\end{equation}
The problem being to choose suitable successive directions of minimisation.

\subsubsection{Optimum direction of minimisation}

In principle, the optimum direction of minimisation  $\delta\M{f}$  could be
derived from the Taylor's expansion:
\begin{equation}
	Q(\M{f}+\delta\M{f})\simeq Q(\M{f})
	+ \sum_k\delta\!f_k \frac{\partial{}Q}{\partial{}f_k}
	+ \frac{1}{2}\sum_{k,l}\delta\!f_k \delta\!f_l
	\frac{\partial^2Q}{\partial{}f_k\partial{}f_l} \,, \label{e:Taylor}
\end{equation}
that is minimised for the step
\begin{equation}
	\delta\M{f} = -\bigl(\nabla\!\nabla\!Q\bigr)^{-1}\mdot\nabla\!Q \, ,
	\label{e:stepQ}
\end{equation}
where $\nabla\!Q$  and  $\nabla\!\nabla\!Q$  are  respectively  the gradient
vector and the Hessian matrix of $Q(\M{f})$:
\begin{displaymath}
	\nabla\!Q_k = \frac{\partial{}Q}{\partial{}f_k} \,,
	\quad\mbox{and}\quad
	\nabla\!\nabla\!Q_{k,l}
		= \frac{\partial^2Q}{\partial{}f_k\partial{}f_l}\,.
\end{displaymath}
The whole difficulty of multi-dimensional minimisation deals with estimating
the inverse of the Hessian matrix, which may generically be too large  to be
computed and stored.  A further difficulty arises when $Q(\M{f})$  is highly
non-quadratic  (e.g.\  in  MEM)  since  the  behaviour  of   $Q(\M{f})$  can
significantly differ from that of its Taylor's expansion.

There exist a number of  multi-dimensional  minimisation  numerical routines
that avoid the direct computation of  the  inverse  of  the  Hessian matrix:
e.g.\ steepest descent, conjugate gradient algorithm, Powell's  method, etc.
\cite{NumericalRecipes}.  For the steepest descent method, the  direction of
minimisation      is      simply      given      by       the      gradient:
$\delta\M{f}_\R{SD}=-\nabla\!Q$.   Other  more  efficient  multi-dimensional
minimisation methods attempt to build information  about  the  Hessian while
deriving  a  more  optimal  direction,  i.e.\  a  better   approximation  of
$-(\nabla\!\nabla\!Q)^{-1}\mdot\nabla\!Q$.       For      instance,      the
conjugate-gradient method builds a series  of  optimum  conjugate directions
$\delta\M{f}_\R{CG}$, each of which is a linear combination  of  the current
gradient and the previous  direction  \cite{NumericalRecipes}.   Among those
improved methods and when the number of parameters is very large, the choice
of  conjugate-gradient  is  driven  by  its  efficiency  both  in  terms  of
convergence rate and memory allocation.

\subsubsection{Accounting for positivity}\label{s:pos1d}

Let us now  examine  the  non-linear  strategy  leading  to  a  minimisation of
$Q(\M{f})$ with the constraint that  $\hat{f}(\eta,h)\ge0$  everywhere.   We
will assume that the basis of functions $\{\hat{e}_k(\eta,h)\}$ is chosen so
that  the   positivity   constraint   is   equivalent   to   enforcing  that
$f_k\ge0;\forall{}k$ (see Appendix~\ref{section:linearInterpolation}  for an
example of such a  basis).

When seeking the appropriate step size given by \Eq{monocp}, it  is possible
to account for positivity by limiting the range of $\lambda^{(n)}$:
\begin{displaymath}
	\M{f}^{(n+1)}\ge 0
	\quad \Leftrightarrow \quad
	-\StackUnder{\R{min}}{\delta\!f^{(n)}_k>0}
	\frac{f^{(n)}_k}{\delta\!f^{(n)}_k}
	\le \lambda^{(n)} \le
	-\StackUnder{\R{max}}{\delta\!f^{(n)}_k<0}
	\frac{f^{(n)}_k}{\delta\!f^{(n)}_k}.
\end{displaymath}
In practice this procedure blocks the  steepest  descent method
long before the right solution is found.  It is in fact  better  to truncate
negative values after each step:
\begin{displaymath}
	f_k^{(n+1)}=\max\{0, f_k^{(n)}+\lambda^{(n)}\,\delta{}f_k^{(n)}\}\,.
\end{displaymath}
Besides,   any   of   these   methods   to    enforce    positivity   breaks
conjugate-gradient minimisation since  this  latter  assumes  that  the true
minimum  of  $Q(\M{f}^{(n)}+\lambda^{(n)}\,\delta\M{f}^{(n)})$   is  reached
while varying $\lambda^{(n)}$.

Thi\'ebaut \& Conan \shortcite{ThiebautConan95}  circumvent  this difficulty
thanks to a reparametrisation  that  enforces  positivity.   Following their
argument, $Q$ is minimised here with respect  to  a  new  set  of parameters
$\M{x}$ such as:
\begin{equation}
	f_k = g(x_k)\,, \quad \mbox{with $g: \R{I\!R} \mapsto \R{I\!R}_+$}\,.
	\label{e:reparam}
\end{equation}
The following various reparametrisations meet these requirements:
\begin{displaymath}
	\begin{array}{lcl}
	f_k = \mathrm{exp}(x_k) &\Rightarrow& g'(x_k)^2 = f_k^2\,, \\
	f_k = x_k^{2n}\quad\mbox{($n$ positive integer)}
		&\Rightarrow& g'(x_k)^2 \propto f_k^{2-1/n}\,. \\
	\end{array}
\end{displaymath}

When $Q(\M{f})$  is  quadratic,  $Q(\M{f}+\lambda\delta\M{f})$  is  a second
order polynomial with respect to $\lambda$, the  minimisation  of  which can
trivially be performed with a very limited number of matrix multiplications.
One  drawback  of  the  reparametrisation  is  that,  since   $g(\cdot)$  is
non-linear, $Q\!\circ\!g(\M{x})$ is necessarily non-quadratic.  In that case
the  exact   minimisation   of   $Q\!\circ\!g(\M{x}+\lambda\delta\M{x})$  --
mandatory in conjugate gradient or Powell's methods  --  requires  many more
matrix  multiplications.   Another  drawback  is  that   the   direction  of
investigation derived by conjugate gradient or Powell's  methods  may  be no
longer optimal requiring many more steps  to  obtain  the  overall solution.
This  latter  point  follows  from  the  fact  that  these  methods  collect
information about the Hessian while taking into account the  previous steps,
whereas for a non-quadratic  functional  this  information  becomes obsolete
very soon since the Hessian (with respect to $\M{x}$) is no  longer constant
\cite{SkillingBryan84}.

Consequently, instead of  varying  the  parameters  $\M{x}$,  we  propose to
derive a step $\delta\M{f}$ for varying $\M{f}$  from  the reparametrisation
that enforces positivity.  Let  $\delta\M{x}$  be  the  chosen  direction of
minimisation    for     $\M{x}$,     the     sought     parameters    reads:
$g(x_k+\lambda\,\delta{}x_k)\simeq{}f_k+\lambda\,\delta{}x_k\,g'(x_k)$.
Identifying   the   right    hand    side    of    this    expression   with
$f_k+\lambda\delta\!f_k$ yields: $\delta\!f_k=\delta{}x_k\,g'(x_k)$.   Using
the steepest descent direction:
\begin{displaymath}
	\delta{}x_k=-\frac{\partial{}Q}{\partial{}x_k}
	= - \frac{\partial{}Q}{\partial{}f_k} g'(x_k)\,.
\end{displaymath}
yielding finally:
\begin{equation}
	\delta\!f_k = - \frac{\partial{}Q}{\partial{}f_k} g'(x_k)^2\,
	= - \frac{\partial{}Q}{\partial{}f_k} f_k^\nu\,,
	\label{e:deltafk}
\end{equation}
with $1\le\nu\le2$ depending on the particular choice of $g(\cdot)$.

\subsubsection{Algorithm for 1D minimisation: positivity at fixed $\mu$}

In Appendix~\ref{s:mem} we show that other authors have derived very similar
optimum direction of minimisation but   in the more   restrictive case of  a
regularisation by $R_\R{MEM}$. Note that our approach is not limited to this
type of penalising function since  positivity is enforced extrinsically.  In
short, the minimisation step is derived from the unifying expression:
\begin{equation}
	\delta\!f_k = -q_k \nabla\!Q_k \,,
	\label{e:step-single}
\end{equation}
where the gradient is scaled by (see Appendix~\ref{s:mem}):
\begin{displaymath}
	q_k = \left\{\begin{array}{cl}
		f_k^\nu\, , & \mbox{this work (with $1\le\nu\le2$)}\,, \\[.7ex]
		f_k \, ,	& \mbox{Richardson-Lucy}\, ,\\[.7ex]
		f_k/\mu \, ,	& \mbox{classical MEM}\, ,\\[1.2ex]
		{\displaystyle\frac{f_k}{\mu + f_k \sum\limits_{i,j}
			\frac{a_{i,j,k}^2}{\R{Var}(\tilde{F}_{i,j})}}}
		\,,	& \mbox{Cornwell-Evans.}\\
	\end{array}\right.
\end{displaymath}

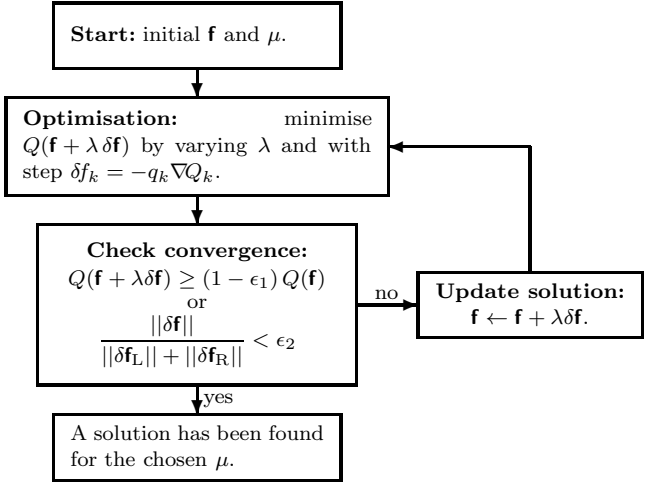
\begin{figure}
\begin{center}
\noindent\unitlength=0.01\columnwidth
\begin{picture}(100,75)
	\thicklines
	\put(7.5,65){\framebox(45,10){\parbox[c]{40\unitlength}{
		{\bf Start:} initial $\M{f}$ and $\mu$.}}}
	\put(30,65){\vector(0,-1){5}}
	 \put(0,45){\framebox(60,15){\parbox[c]{55\unitlength}{
		{\bf Optimisation:} minimise
		\mbox{$Q(\M{f}+\lambda\,\delta\M{f})$} by varying $\lambda$
		and with step \mbox{$\delta\!f_k=-q_k\nabla\!Q_k$}.}}}
	\put(30,45){\vector(0,-1){5}}
	\put(5,15){\framebox(50,25){\shortstack{{\bf Check convergence:}\\
		$Q(\M{f}+\lambda\delta\M{f})\ge(1-\epsilon_1)\,Q(\M{f})$\\
		or\\
		${\displaystyle\frac{||\delta\M{f}||}
		{||\delta\M{f}_\R{L}||+||\delta\M{f}_\R{R}||}< \epsilon_2}$}}}
	\put(30,15){\vector(0,-1){5}}
	\put(31,12){yes}
	\put(55,27.5){\vector(1,0){10}}
	\put(58,28.5){no}
	\put(65,22.5){\framebox(35,10){\shortstack{
		{\bf Update solution:}\\
		$\M{f}\leftarrow\M{f}+\lambda\delta\M{f}$.}}}
	\put(82.5,32.5){\line(0,1){20}}
	\put(82.5,52.5){\vector(-1,0){22.5}}
	\put(7.5,0){\framebox(45,10){\parbox[c]{40\unitlength}{A solution
		has been found for the chosen $\mu$.}}}
\end{picture}
\end{center}
\caption{Synopsis     of     the      single      direction     minimisation
algorithm.\label{f:alg-single}.        Here
$\epsilon_1\simeq10^{-8}$ and $\epsilon_2\simeq10^{-5}$.}
\end{figure}

The   scheme   of   the   1D-optimisation   algorithm   is   illustrated  in
Fig.~\ref{f:alg-single}.   Iterations  are  stopped  when  the  decrement in
$Q(\M{f})$ becomes negligible, i.e.\ when:
\begin{displaymath}
	\Abs{Q(\M{f}+\lambda\delta\M{f})-Q(\M{f})}
	\le \epsilon \, \Abs{Q(\M{f})} \, ,
\end{displaymath}
where $\epsilon>0$ is a small number which should not  be  smaller  than the
square  root  of  the  machine  precision   \cite{NumericalRecipes}.    Lucy
\shortcite{Lucy94} has suggested another stop criterion based  on  the value
of the ratio:
\begin{displaymath}
	\rho = \Abs{\Abs{\delta\M{f}}} \, /  \,
	(\Abs{\Abs{\delta\M{f}_\R{L}}}+\Abs{\Abs{\delta\M{f}_\R{R}}})\,,
\end{displaymath}
where $\delta\M{f}_\R{L}$ and $\delta\M{f}_\R{R}$ are  the  directions which
minimise the likelihood and the regularisation terms:
\begin{displaymath}
	\delta\M{f}_\R{L}=-\M{q} \times \nabla\!L \quad\mbox{and}\quad
	\delta\M{f}_\R{R}=-\mu \M{q} \times \nabla\!R \, , 
\end{displaymath}
where $\times$ denotes element-wise product (in other  words  $\M{q}$ stands
loosely for $\R{Diag}(q_{1},\ldots,q_{n})$).
In practice  and  regardless  of  the  particular  choice  for  $\M{q}$, the
algorithm makes no significant progress  when  $\rho$  becomes  smaller than
$10^{-5}$.

\subsubsection{Performance issues}

During   the   tests,   it was found   that   conjugate-gradient   method  with
reparametrisation   and   iterative   methods   with   direction   given  by
Eq.~(\ref{e:step-single}) require roughly the same number of steps (one step
involving minimisation along a  new  direction  of  minimisation).   Yet the
non-linear reparametrisation required  to  enforce  positivity  in conjugate
gradient method prevents interpolation and in effect spends  much  more time
(a factor of 10 to 20) to  perform  line  minimisation.   Besides,  when the
current estimate is far from the solution, minimisation  direction following
our prescription~(\ref{e:deltafk}) or that of classical MEM  or  Lucy spends
fewer steps than that of Cornwell \& Evans to bring  $\M{f}$  near  the true
solution.  When the current estimate is sufficiently close to  the solution,
Cornwell \& Evans's method spends half as many steps as the other methods to
reach   the   solution.    The   best   compromise   is   to    start   with
$\delta\!f_k\propto-f_k\nabla\!Q_k$,   then   after   some   iterations  use
$\delta\M{f}=\delta\M{f}_\R{CE}$.  As a rule of thumb,  for low
signal-to-noise ratios  ($\SNR\sim5$)  about  as  many  steps  as  number of
parameters are required, for high signal-to-noise ratios ($\SNR\ge30$) fewer
steps are needed (up to 10 times less).  It remains that with  these methods
trial and error iterations are required to find  the  appropriate  value for
$\mu$.   The  different  implementations  are  illustrated  and  compared in
\Fig{fig1} and \Fip{fig2} as described in section~(\ref{sect:sim}).

Accounting for positivity in multi-dimensional optimisation therefore 
leads to a modified steepest descent algorithm for which the current 
gradient is locally rescaled.  A faster 
convergence is achieved when some information from the Hessian is 
extracted appropriately.  Yet, the above described algorithm assumes 
that optimisation is performed with a fixed value of the Lagrange 
parameter $\mu$.  Let us now turn to a more general minimisation along 
several direction which allow $\mu$ to be adjusted on the fly during 
the minimisation.

\subsection{Minimisation along several directions }
\label{s:multi-dir}

\subsubsection{Skilling \& Bryan method revisited}

In the context of  maximum  entropy  image  restoration,  Skilling  \& Bryan
\shortcite{SkillingBryan84} (hereinafter SB) have proposed a powerful method
which is both efficient in optimising  a  non-linear  problem  with  a great
number  of  parameters  and  able  to  vary  automatically  the   weight  of
regularisation so that  the  sought  solution  satisfies $L(\M{f})=N_\R{e}$.
Here their approach is further generalise  to any  penalising  function.   In
short, SB derive their method from the following remarks:
\begin{enumerate}
\item To account for positivity, they suggest  an appropriate ``metric'' (or
      rescaling)  which  is  equivalent to   multiplying  each  minimisation
      direction by $\M{q}=\M{f}$.
\item The regularisation weight $\mu$ is adjusted at each iteration  to meet
      the constraint $L(\M{f})=N_\R{e}$.  Therefore,  instead  of minimising
      along the single direction  $-\M{q}\times(\nabla\!L+\mu\nabla\!R)$, at
      least  two  directions  are  considered:  $-\M{q}\times\nabla\!L$  and
      $-\M{q}\times\nabla\!R$.
\item Since the Hessian is not constant -- at least since  $\mu$  is allowed
      to vary, no information is carried from the previous iterations.  This
      clearly excludes conjugate-gradient or  similar  optimisation methods,
      but favours non-quadratic penalising functions for  which  the Hessian
      is not assumed to be constant.
\item If the whole  Hessian  cannot  be  computed,  it  can  nevertheless be
      applied to any vector $\M{e}$ of same  size  as  $\M{f}$  in  a finite
      number  of  operations,  e.g.\  two  matrix  multiplications  for  the
      likelihood         term:         $\nabla\!\nabla\!L\mdot\M{e}        =
      2\,\T{\M{a}}\mdot(\M{a}\mdot\M{e})$   (where,   for   the    sake   of
      simplicity, the diagonal weighting matrix was omitted here).  They
      illustrate how this provide a means to include some knowledge from the
      local Hessian while seeking the optimum minimisation direction.
\end{enumerate}

\subsubsection{Local minimisation sub-space}

In order to adjust the  regularisation  weight,  at  least  two simultaneous
directions  of  minimisation  should  be  used:  $-\M{q}\times\nabla\!L$ and
$-\M{q}\times\nabla\!R$.   Furthermore,  the  local  Hessian  provides other
directions of minimisation to increase the convergence  rate.   Using matrix
notation, the Taylor expansion of $Q(\M{f})$ for two simultaneous directions
$\delta\M{f}_1$ and $\delta\M{f}_2$ reads:
\begin{eqnarray}
	Q(\M{f}\!+\!\delta\M{f}_1\!+\!\delta\M{f}_2)
	&\!\!\simeq\!\!& Q(\M{f})
	   + \delta\T{\M{f}_1}\!\mdot\!\nabla\!Q
           + \frac{1}{2}
	     \delta\T{\M{f}_1}\!\mdot\nabla\!\nabla\!Q\mdot\delta\M{f}_1
	\nonumber \\
	&& + \delta\T{\M{f}_2}\!\mdot
	     [\nabla\!Q+\nabla\!\nabla\!Q\mdot\delta\M{f}_1]
           + \frac{1}{2}
	     \delta\T{\M{f}_2}\!\mdot\nabla\!\nabla\!Q\mdot\delta\M{f}_2\,,
	\nonumber
\end{eqnarray}
where the Hessian and gradient are  evaluated  at  $\M{f}$.   Given  a first
direction $\delta\M{f}_1$, the optimal choice for a second direction is:
\begin{displaymath}
	\delta\M{f}_{2,\R{opt}} = - (\nabla\!\nabla\!Q)^{-1}\mdot
	(\nabla\!Q+\nabla\!\nabla\!Q\mdot\delta\M{f}_1)\,.
\end{displaymath}
In MEM, recall that positivity is enforced explicitly by  the regularisation
penalty  function  while  efficient  minimisation   methods   rely   on  the
approximation of $(\nabla\!\nabla\!Q)^{-1}$  by  a  scaling  vector $\M{q}$.
The optimum first two directions then become in MEM:
\begin{displaymath}
	\delta\M{f}_{1,\R{opt}} \simeq - \M{q} \times \nabla\!Q\,,
	\quad \mbox{and} \quad
	\delta\M{f}_{2,\R{opt}} \simeq
	- \M{q} \times (\nabla\!Q+\nabla\!\nabla\!Q\mdot\delta\M{f}_1)
	\,,
\end{displaymath}
Since   the   first   term   in   the   right   hand   side   expression  of
$\delta\M{f}_{2,\R{opt}}$ is $\delta\M{f}_{1,\R{opt}}$, the two near optimum
directions sought are finally:
\begin{equation}
	\delta\M{f}_1 = - \M{q} \times \nabla\!Q \,,
	\quad \mbox{and} \quad
	\delta\M{f}_2 =
	- \M{q} \times (\nabla\!\nabla\!Q\mdot\delta\M{f}_1) \, .
	\label{e:step-multi-MEM-2}
\end{equation}
Similar  considerations   yield here further  possible  directions:
\begin{equation}
	\delta\M{f}_n =
	- \M{q} \times (\nabla\!\nabla\!Q\mdot\delta\M{f}_{n-1}) \, .
	\label{e:step-multi-MEM-3}
\end{equation}
If the rescaling, $\M{q}$, provides too good an approximation of the inverse
of the Hessian, then $\delta\M{f}_1$  and  $\delta\M{f}_2$  would  almost be
identical (i.e.\ antiparallel); hence using  only  one  is  sufficient.   In
other words, since  the  local  Hessian  is  accounted  for  by  the  use of
additional   directions   of   minimisation,   there   is   no   need   that
$\R{Diag}(\M{q})$      be      an       accurate       approximation      of
$(\nabla\!\nabla\!Q)^{-1}$.      The     crude     rescaling     given    by
Eq.~(\ref{e:delta-f-MEM})   is   therefore    sufficient,    i.e.\   taking:
$\M{q}=\M{f}$.  This definition of  $\M{q}$  has  the  further  advantage to
warrant positive values of $\M{f}$ and does not depend on  the  actual value
of $\mu$ (which is obviously not the case for the Hessian).

If no term in $Q(\M{f})$ enforces positivity, it was shown earlier  that the
reparametrisation~(\ref{e:reparam}) would.   From  the  Taylor  expansion of
$Q(\M{x})$, the first two steepest descent directions  with  respect  to the
parameters $\M{x}$ are given by:
\begin{eqnarray}
	\delta{}x_{1,k}
	&\!=\!& - \frac{\partial{}Q}{\partial{}x_k} 
	\!=\! - g'(x_k) \nabla\!Q_k \,,  \label{e:cpdir} \\
	\delta{}x_{2,k}
	&\!=\!& - \sum_l \frac{\partial^2Q}{\partial{}x_k\partial{}x_l}
		         \frac{\partial{}Q}{\partial{}x_l}
	\!=\! - g'(x_k) \sum_l \nabla\!\nabla\!Q_{k,l}
	          g'(x_l) \delta{}x_{1,l} \,.\nonumber
\end{eqnarray}
Since   $\delta{}f_k\simeq{}g'(x_k)\delta{}x_k$,   the   near   two  optimum
directions of minimisation for the parameters $\M{f}$ read:
\begin{eqnarray}
	\delta{}f_{1,k} &=& - g'(x_k)^2 \nabla\!Q_k \,, \\
	\delta{}f_{2,k} &=& - g'(x_k)^2
		\sum_l \nabla\!\nabla\!Q_{k,l} \delta{}f_{1,l} \,,
\end{eqnarray}
which    are     incidentally     identical     to     those     given    by
Equation~(\ref{e:step-multi-MEM-2}), provided that $q_k=g'(x_k)^2$.

For all the regularisation penalising functions considered here, clearly the
best choice is to use directions given by the Hessian applied  to \Eq{cpdir}
when other directions of minimisation than those related to the gradient are
considered.  Since $\mu$ can vary, the Hessians of $L$ and  $R$  have  to be
applied separately.  At each step, the minimisation  is  therefore performed
in the $n=3\times 2=6$ dimensional sub-space defined by:
\begin{equation}
\begin{array}{ll}
	\delta\M{f}_1 = -\M{q} \times \nabla\!L \,,&
	\delta\M{f}_2 = -\M{q} \times \nabla\!R \,,\\
	\delta\M{f}_3 = -\M{q} \times
		(\nabla\!\nabla\!L\mdot\delta\M{f}_1) \,,&
	\delta\M{f}_4 = -\M{q} \times
		(\nabla\!\nabla\!L\mdot\delta\M{f}_2)\,,\\
	\delta\M{f}_5 = -\M{q} \times
		(\nabla\!\nabla\!R\mdot\delta\M{f}_1) \,,&
	\delta\M{f}_6 = -\M{q} \times
		(\nabla\!\nabla\!R\mdot\delta\M{f}_2)\,,\\
\end{array}
\end{equation}
where
\begin{equation}
	q_k = f_k^\nu \quad \mbox{with} \quad 1\leq\nu\leq2 \,.
\end{equation}
When $\nu=1$, $\M{q}$ is the same metric  as  that  introduced  by  SB while
relying  on  other  arguments.   Depending  on  the  actual  expression  for
$\nabla\!\nabla\!R$ (and in particular in  MEM  with  a  constant  prior), a
smaller number of directions need be explored (for  instance,  SB  used only
$n=3$     simultaneous     directions     since      when     $\M{q}=\M{f}$,
$\nabla\!\nabla\!R=1/\M{f}$   so   $\delta\M{f}_5    =    \delta\M{f}_1   $,
$\delta\M{f}_6 = \delta\M{f}_2$;  they  also  use  a  linear  combination of
$\delta\M{f}_3$ and $\delta\M{f}_4$).  In this $n$-dimensional  sub-space, a
simple        second        order         Taylor         expansion        of
$Q(\M{f}+\sum_{i=1}^n\lambda_i\delta\M{f}_i)$, shows that the optimum set of
weights sought, $\{\lambda_1,...,\lambda_n\}$, is given by  the  solution of
the $n$ linear equations parametrised by $\mu$ and given by:
\begin{equation}
	\sum_{j=1}^n \lambda_j \delta\T{\M{f}}_j \mdot
		(\nabla\!\nabla\!L+\mu\nabla\!\nabla\!R) \mdot
		\delta\M{f}_i
	= - \delta\T{\M{f}}_i \mdot (\nabla\!L+\mu\nabla\!R)\,.
	\label{e:multi-lambda}
\end{equation}
Now in that sub-space, the optimisation may  be  ill-conditioned  (i.e.\ the
set of linear equations are linearly dependent  in  a  numerical  sense). In
order  to   deal   with   this   degeneracy   truncated   SVD  decomposition
\cite{NumericalRecipes} is used to find  a  set  of  numerically independent
directions.  In practice, the rank of the 6 linear equations  varies  from 2
(very far from the  solution  or  when  convergence  is  almost  reached) to
typically 5 or 6.  This method turns out to be much easier to implement than
the bi-diagonalisation suggested by SB.

\subsubsection{On the fly derivation of the regularisation weight}

At each iteration a strategy similar to  that  of  SB  was  adopted  here to
update the value of $\mu$:
\begin{enumerate}
\item  $L_\R{min}$ and $L_\R{max}$  are  the  values  of the
      likelihood term in the sub-space in the  limits  $\mu\rightarrow0$ and
      $\mu\rightarrow\infty$ respectively.  The corresponding solutions give
      what  we  call  the  maximum  likelihood  solution  and   the  maximum
      regularised solution in the sub-space.
\item If $L_\R{max}<N_\R{e}$ the maximum regularised  solution corresponding
      to  $L_\R{max}$  is  adopted  to  proceed  to   the   next  iteration.
      Otherwise, in order to avoid relaxing the regularisation and following
      SB, a modest reachable goal is fixed:
\begin{displaymath}
	L_\R{aim}=\R{max}\{N_\R{e},(1-\alpha)L_\R{prev}+\alpha{}L_\R{min}\}
\end{displaymath}
      where $L_\R{prev}$ is the likelihood value at the end of  the previous
      iteration, while $0\!<\!\alpha\!<\!1$  (say  $\alpha=2/3$).   A simple
      bi-section method is applied to seek the value of $\mu$ for  which the
      solution of Eq.~(\ref{e:multi-lambda}) yields $L=L_\R{aim}$.
\end{enumerate}
Following this scheme, the algorithm varies the value of  $\mu$  so  that at
each iteration the likelihood is reduced until it reaches its  target value;
then the regularisation  term  is  minimised  while  the  likelihood remains
constant.

As a stop criteria, a measure of  the  statistical  discrepancy  between two
successive iterations:
\begin{displaymath}
	\sum_k f^{(n)}_k\Abs{f^{(n+1)}_k-f^{(n)}_k}
	\quad/\quad\sum_k f^{(n)}_k\,,
\end{displaymath}
is computed.  In practice, in order to avoid  over-regularisation, $N_\R{e}$
is taken to be $N_\R{data}-\sqrt{2N_\R{data}}$.  The corresponding scheme of
the   $n$-dimensional    optimisation    algorithm    is    illustrated   in
Fig.~\ref{f:alg-multi}.

\newsavebox{\myBoxA}
\newsavebox{\myBoxB}
\newsavebox{\myBoxC}
\newsavebox{\myBoxD}
\newsavebox{\myBoxE}
\newsavebox{\myBoxF}
\newsavebox{\myBoxG}
\newsavebox{\myBoxH}
\newsavebox{\myBoxI}
\newsavebox{\myBoxJ}
\sbox{\myBoxA}{\parbox{7.8cm}{{\bf Initialisation Step:}
\begin{list}{}{\settowidth{\labelwidth}{(1)}
\labelsep1ex\leftmargin=\labelwidth
\advance\leftmargin by 2\labelsep
\itemindent0pt
\listparindent0pt
\itemsep1ex
\parsep1ex
\topsep0ex
\parskip1ex
}
\item[(1)] Compute the $n$ directions of minimisation:\\[1ex]
$\hbox{}\quad\begin{array}{ll}
	\delta\M{f}_1 = \M{q} \!\times\! \nabla\!L,&
	\delta\M{f}_2 = \M{q} \!\times\! \nabla\!R,\\
	\delta\M{f}_3 = \M{q} \!\times\!
		(\nabla\!\nabla\!L\mdot\delta\M{f}_1),&
	\delta\M{f}_4 = \M{q} \!\times\!
		(\nabla\!\nabla\!L\mdot\delta\M{f}_2),\\
	\delta\M{f}_5 = \M{q} \!\times\!
		(\nabla\!\nabla\!R\mdot\delta\M{f}_1),&
	\delta\M{f}_6 = \M{q} \!\times\!
		(\nabla\!\nabla\!R\mdot\delta\M{f}_2),\\
	...\\
\end{array}$\\[1ex]
and the coefficients:\\[1ex]
$\hbox{}\quad{}A^L_{i,j}=\delta\T{\M{f}}_j\mdot\nabla\!\nabla\!L
\mdot\delta\M{f}_i\,,\quad{}B^L_i=-\delta\T{\M{f}}_i\mdot\nabla\!L$,\\[1ex]
$\hbox{}\quad{}A^R_{i,j}=\delta\T{\M{f}}_j\mdot\nabla\!\nabla\!R
\mdot\delta\M{f}_i\,,\quad{}B^R_i=-\delta\T{\M{f}}_i\mdot\nabla\!R;$\\[1ex]
define:\\[1ex]
$\hbox{}\quad{}L_\mu\equiv{}L(\M{f}+\sum\limits_{i=1}^n
\lambda_i^\mu\delta\M{f}_i)$\\[1ex]
where $\{\lambda_1^\mu,...,\lambda_n^\mu\}$ minimises
$Q(\M{f}+\sum_i\lambda_i^\mu\delta\M{f}_i)$ and is
derived by solving by truncated SVD method the system:\\[1ex]
$\hbox{}\quad{}\sum\limits_{j=1}^n(A^L_{i,j}+\mu{}A^R_{i,j})
\lambda_j^\mu=B^L_i+\mu{}B^R_i$.
\\[1ex]
\item[(2)] Set: $\begin{array}[t]{l@{=}l}
L_\R{min} & \lim\limits_{\mu\rightarrow0}L_\mu\,,\quad \mu_\R{min}=0\,,\\
L_\R{max} & \lim\limits_{\mu\rightarrow\infty}L_\mu\,,\\
L_\R{aim} & \max\{N_\R{e},(1-\alpha)L(\M{f})+\alpha{}L_\R{min}\}\,,\\
\end{array}$
\\[1ex]
unset $\mu_\R{max}$.
\end{list}
}}
\sbox{\myBoxB}{$L_\R{max}\leq{}L_\R{aim}$}
\sbox{\myBoxC}{$\lambda_j=\lim\limits_{\mu\rightarrow\infty}\lambda_j^\mu$}
\sbox{\myBoxD}{$L_\mu\simeq{}L_\R{aim}$}
\sbox{\myBoxE}{$\lambda_j=\lambda_j^\mu$}
\sbox{\myBoxF}{$L_\mu>L_\R{aim}$}
\sbox{\myBoxG}{$\begin{array}{l@{\leftarrow}l}
	\mu_\R{max} & \mu\\
	\mu         & (\mu+\mu_\R{min})/2\\
	\end{array}$}
\sbox{\myBoxH}{$\begin{array}{l@{\leftarrow}l}
	\mu_\R{min} & \mu\\
	\mu         & \left\{\!\!\!\begin{array}{ll}
	(\mu+\mu_\R{max})/2 & \mbox{if $\mu_\R{max}$ is set}
	\!\!\!\!\!\!\hbox{}\\
	2 \mu & \mbox{otherwise}\\
	\end{array}\right.\\
	\end{array}$}
\sbox{\myBoxI}{\parbox{7.8cm}{{\bf Update solution:}
	$\M{f}\leftarrow\M{f}+\sum\limits_{i=1}^n\lambda_i\delta\M{f}_i$\\
	{\bf Enforce positivity:}
	$f_k\leftarrow\left\{\begin{array}{ll}
	 f_k & \mbox{if $f_k > f_\R{min}$}\!\!\!\!\!\!\!\!\!\!\hbox{}\\
	 f_\R{min} & \mbox{otherwise}\\
	 \end{array}\right.$}}
\sbox{\myBoxJ}{$\displaystyle\frac{\sum_k f_k\left\vert\sum_i\lambda_i
	\delta\!f_{i,k}\right\vert}{\sum_k f_k} \!\leq\! \epsilon$}
\begin{figure*}
\begin{center}
\setlength{\unitlength}{0.00087500in}%
\begin{picture}(7584,5109)(214,-4483)
\thicklines
\put(5041,-3976){\line( 4, 5){180}}
\put(5221,-3751){\line( 1, 0){1350}}
\put(6571,-3751){\line( 4,-5){180}}
\put(6751,-3976){\line(-4,-5){180}}
\put(6571,-4201){\line(-1, 0){1350}}
\put(5221,-4201){\line(-4, 5){180}}
\put(6706,-241){\line( 1, 0){1035}}
\put(4366,344){\line( 1, 1){180}}
\put(4546,524){\line( 1, 0){675}}
\put(5221,524){\line( 1,-1){180}}
\put(5401,344){\line(-1,-1){180}}
\put(5221,164){\line(-1, 0){675}}
\put(4546,164){\line(-1, 1){180}}
\put(4366,-241){\line( 1, 1){180}}
\put(4546,-61){\line( 1, 0){675}}
\put(5221,-61){\line( 1,-1){180}}
\put(5401,-241){\line(-1,-1){180}}
\put(5221,-421){\line(-1, 0){675}}
\put(4546,-421){\line(-1, 1){180}}
\put(4366,-826){\line( 1, 1){180}}
\put(4546,-646){\line( 1, 0){675}}
\put(5221,-646){\line( 1,-1){180}}
\put(5401,-826){\line(-1,-1){180}}
\put(5221,-1006){\line(-1, 0){675}}
\put(4546,-1006){\line(-1, 1){180}}
\put(4906,164){\vector( 0,-1){225}}
\put(4906,-421){\vector( 0,-1){225}}
\put(7291,-826){\line( 1, 0){225}}
\put(7516,-826){\line( 0,-1){1350}}
\put(7516,-2176){\line(-1, 0){3375}}
\put(4141,-2176){\line( 0, 1){1350}}
\put(4141,-826){\vector( 1, 0){225}}
\put(4906,-1006){\vector( 0,-1){225}}
\put(5401,-826){\vector( 1, 0){315}}
\put(5401,-241){\vector( 1, 0){315}}
\put(5401,344){\vector( 1, 0){315}}
\put(7111,-1591){\line( 1, 0){405}}
\put(3916,344){\vector( 1, 0){450}}
\put(5041,-3976){\vector(-1, 0){1125}}
\put(5851,-3526){\vector( 0,-1){225}}
\put(6751,-3976){\vector( 1, 0){450}}
\put(6706,344){\line( 1, 0){1035}}
\put(7741,344){\line( 0,-1){2745}}
\put(7741,-2401){\line(-1, 0){1890}}
\put(5851,-2401){\vector( 0,-1){225}}
\put(7201,-4066){\framebox(450,225){{\bf end}}}
\put(4996, 29){no}
\put(4996,-556){no}
\put(4996,-1141){no}
\put(5446,-196){yes}
\put(5446,389){yes}
\put(5446,-781){yes}
\put(4816,-3931){no}
\put(6841,-3931){yes}
\put(226,-4471){\framebox(3690,5085){\usebox{\myBoxA}}}
\put(4456,322){\usebox{\myBoxB}}
\put(5716,164){\framebox(990,360){\usebox{\myBoxC}}}
\put(4501,-263){\usebox{\myBoxD}}
\put(5716,-421){\framebox(990,360){\usebox{\myBoxE}}}
\put(4501,-848){\usebox{\myBoxF}}
\put(5716,-1051){\framebox(1575,450){\usebox{\myBoxG}}}
\put(4366,-1951){\framebox(2745,720){\usebox{\myBoxH}}}
\put(4141,-3526){\framebox(3645,900){\usebox{\myBoxI}}}
\put(5176,-4021){\usebox{\myBoxJ}}
\end{picture}
\end{center}
\caption{Synopsis of the multiple  directions  multidimensional minimisation
algorithm with adjustment of the regularisation weight.  In this algorithm,
$f_\R{min}>0$  is  a  small  threshold  used  to   avoid   negative  values,
$0\leq\epsilon\ll1$    is     a     small     value     used     to    check
convergence.\label{f:alg-multi}}
\end{figure*}
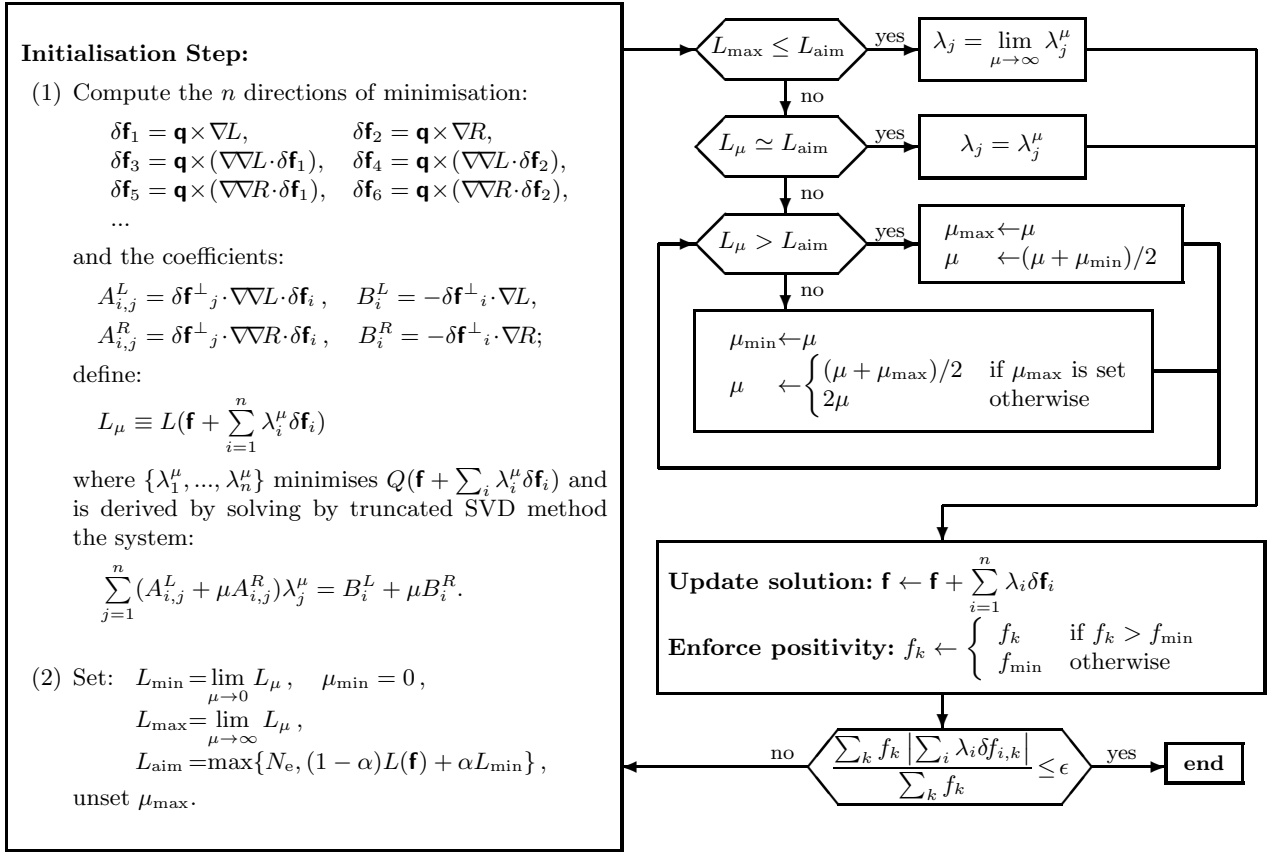

\subsubsection{Performance \& assessments}

The optimisation of $Q(\M{f})$ in a multi-dimensional sub-space  yields many
practical advantages: (i) it provides  faster  convergence  rates  (about 10
times less overall iterations and even much  less  when  accounting  for the
number of inversions required to derive the regularisation weight)  and less
overall {\sc cpu} time  in  spite  of  the  numerous  matrix multiplications
involved to compute the $n$ directions of minimisation and  their  images by
the Hessians.  (ii) It yields a more robust  algorithm  because  it  is less
sensitive to local minima and also because the routine requires less tuning.
(iii) Since $\mu$ varies between iterations and since the  local  Hessian is
always re-estimated, the solution can be modified on the fly, e.g.\ rescaled, 
without perturbing the convergence.   Hence
the normalisation is no more an issue.

This algorithm presents the  following  set  of  improvements  over  that of
Skilling \& Bryan: (i) a more general penalising functions  than  entropy is
considered  (e.g.\  entropy  with  floating  prior  or  quadratic penalising
function) which yield a different {\em metric} derived heuristically.   This
yields almost the same optimisation sub-space but from a different approach;
(ii) truncated SVD is implemented to avoid ill-conditioned problems  in this
minimisation sub-space.

\section{Simulations}\label{sect:sim}

\begin{figure*}
\centerline{\psfig{file=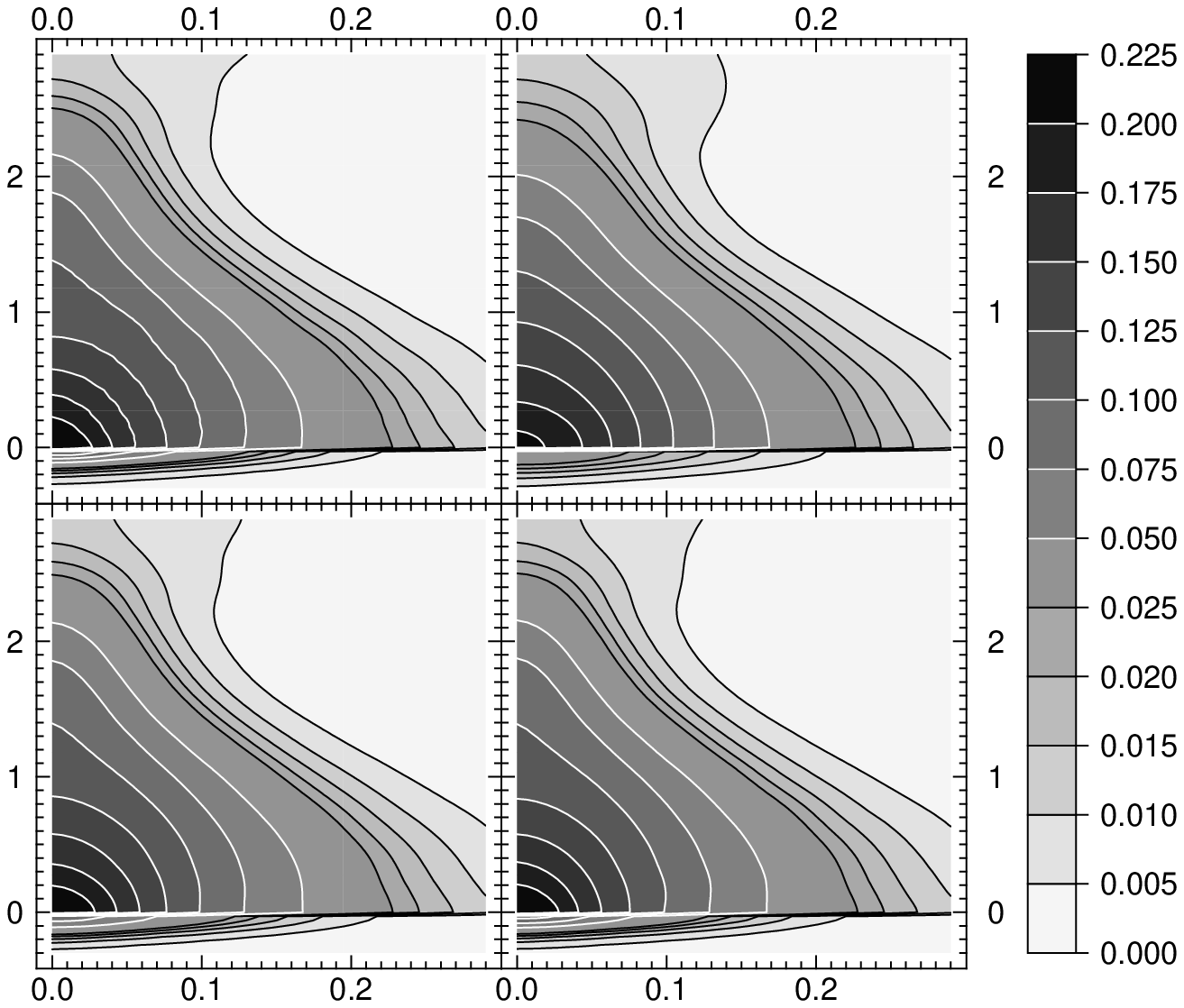,width=1.5\columnwidth}}
\caption{  (from   top-left   to   bottom-right):   (1)   true  distribution
(approximately that of a  disk  with  Toomre  parameter  $Q=1.25$)  and mean
restoration of $\hat{f}(\eta,h)$ out of 40 itterations for a SNR  of  5 (2),
30 (3) and 100 (4).   Abscissa  is  normalised  specific  energy  $\eta$ and
ordinate is specific angular momentum $h$. Note that the isocontours are not
sampled uniformly in order to display more acurately counter rotating stars.
\label{f:simu-f-avg}}
\end{figure*}

\begin{figure*}
\centerline{\psfig{file=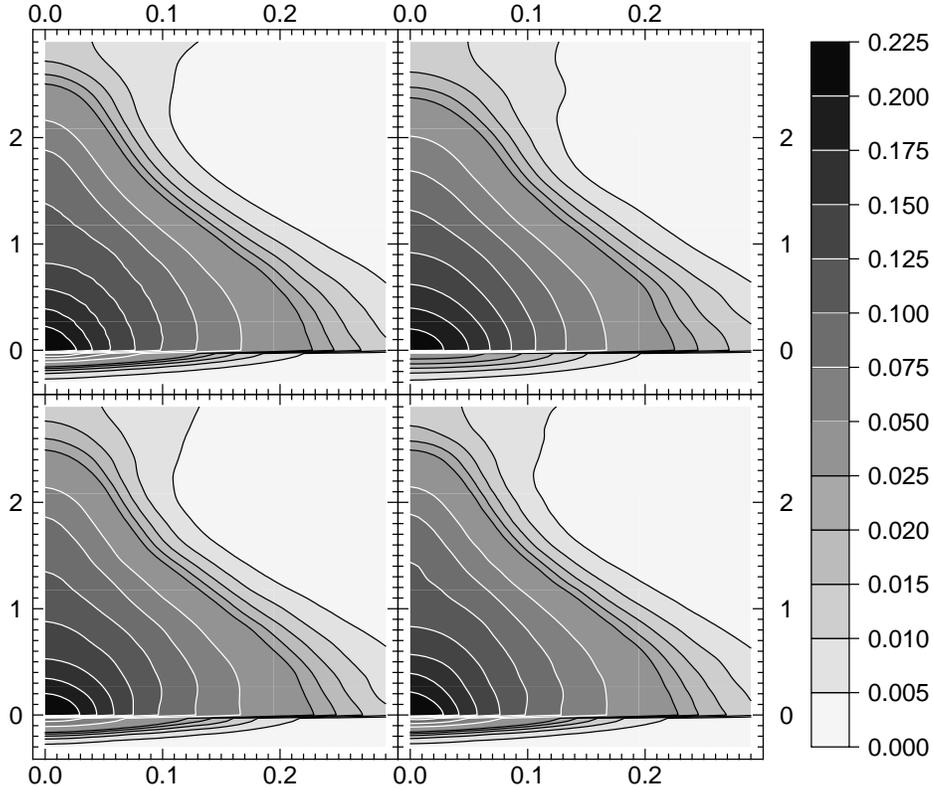,width=1.5\columnwidth}}
\caption{true distribution   and one  sample  for each  SNR out  of  the  40
restorations                carried      out,       displayed     as      in
Fig.~\protect\ref{f:simu-f-avg}. Abscissa  is    normalised  specific energy
$\eta$ and  ordinate  is  specific  angular  momentum $h$. Comparison   with
Fig.~\protect\ref{f:simu-f-avg} shows that the inversion is successfull both
statistically and on a per sample basis.
\label{f:simu-f-sample}}
\end{figure*}

\begin{figure}
\noindent\centerline{\psfig{file=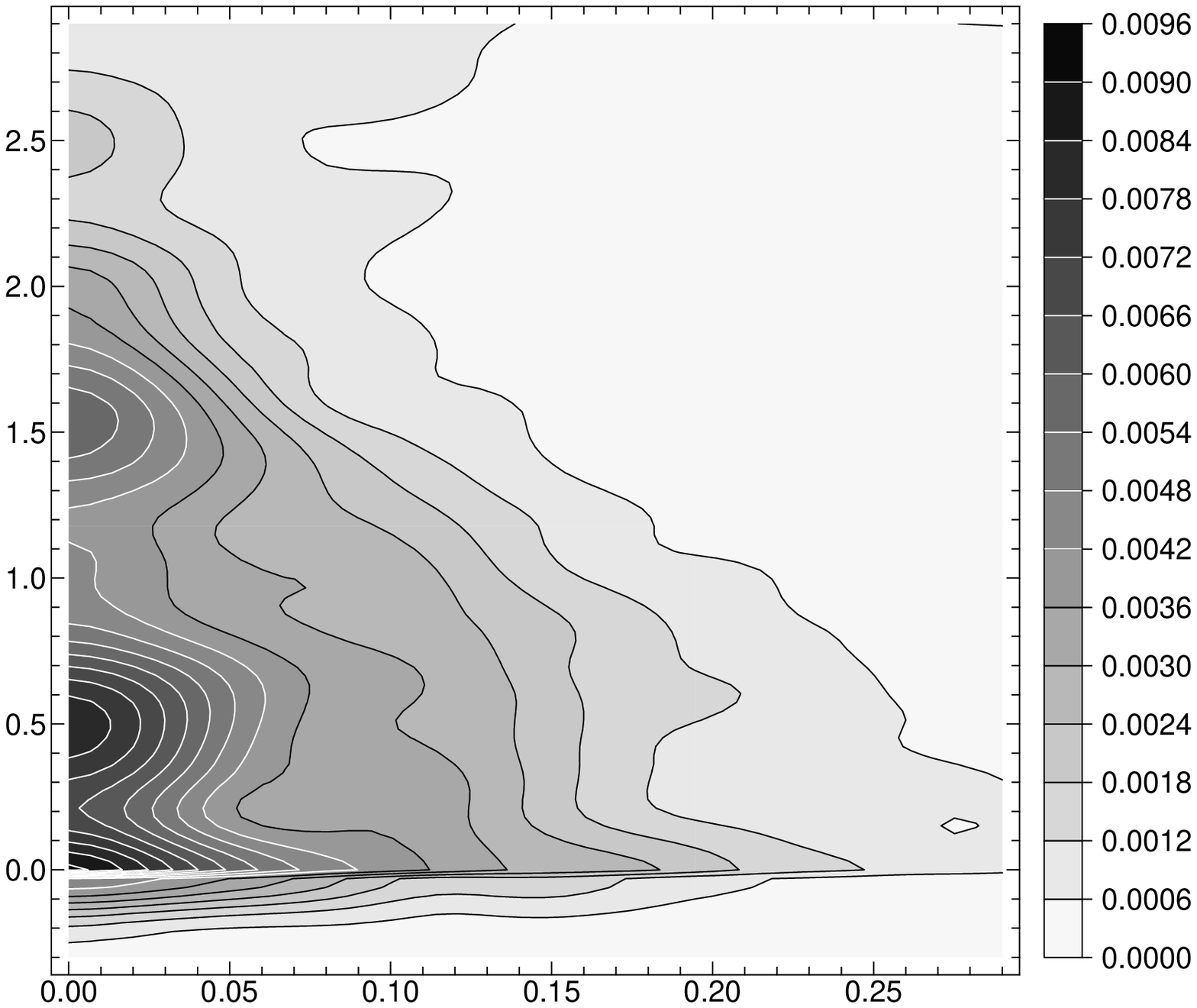,width=\columnwidth}}\\
\noindent\centerline{\psfig{file=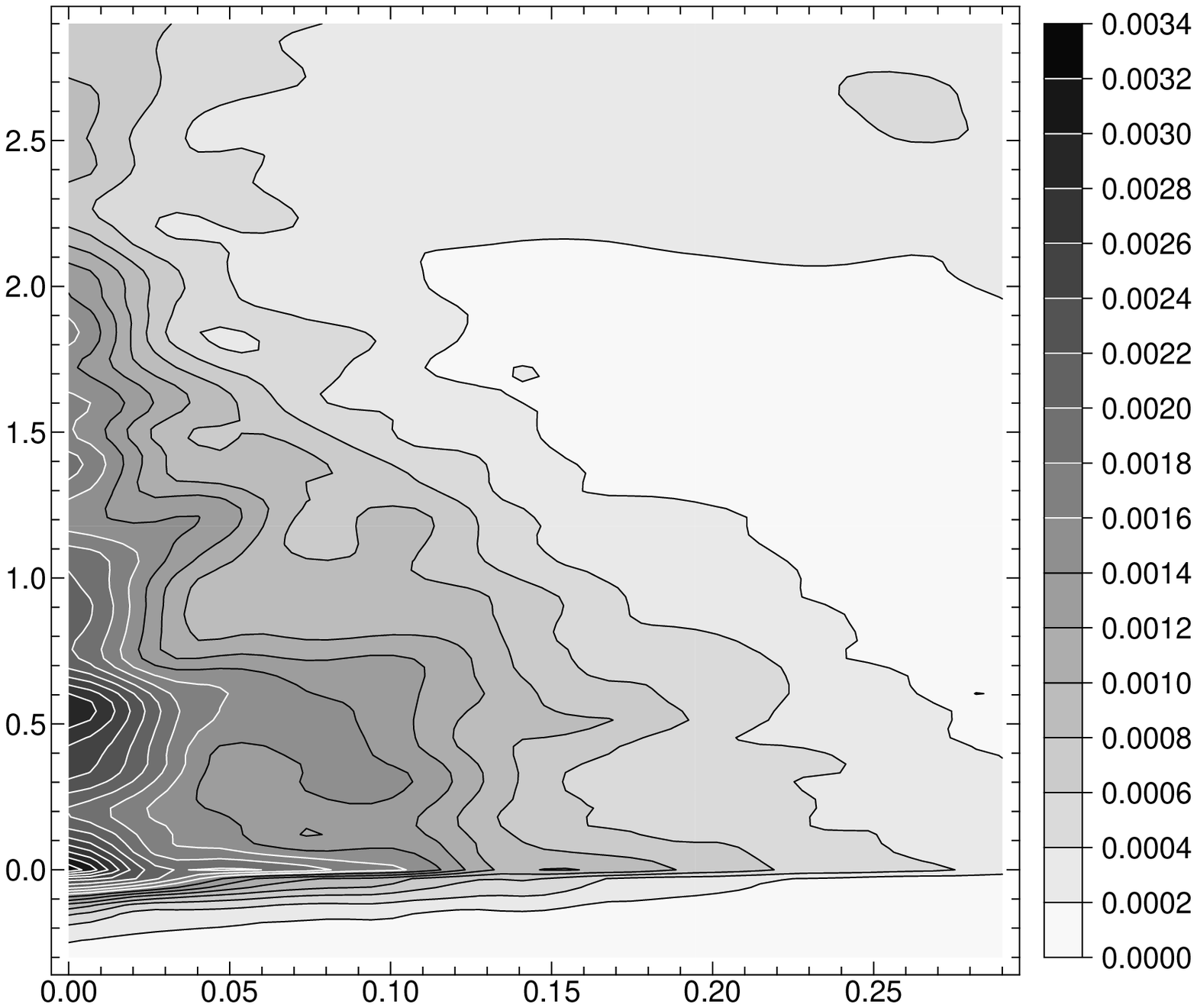,width=\columnwidth}}\\
\noindent\centerline{\psfig{file=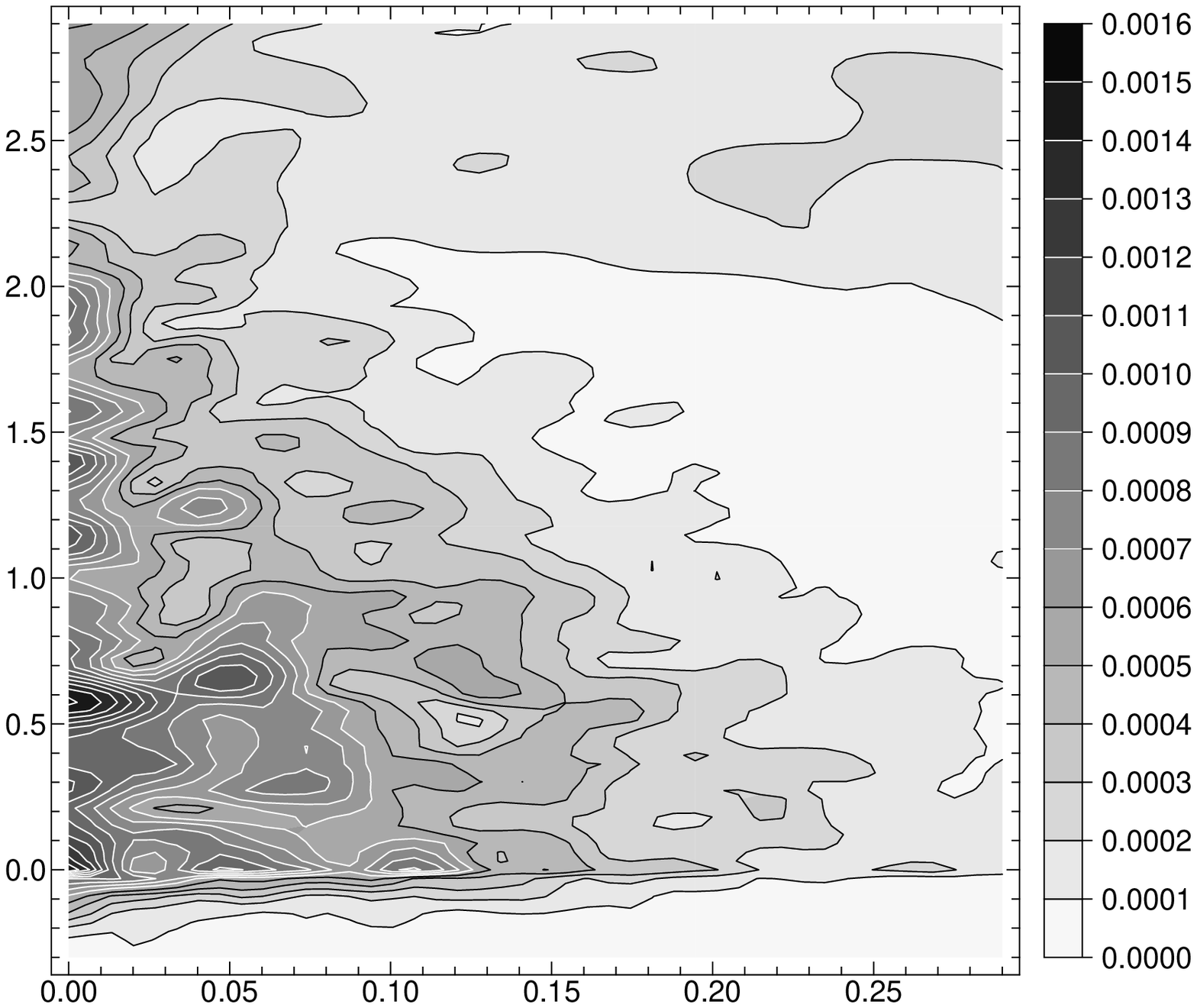,width=\columnwidth}}
\caption{Standard deviation for each SNR out  of the 40 restorations carried
out displayed as in  Fig.~\protect\ref{f:simu-f-avg}.   From top to  bottom:
$\SNR=5$, $\SNR=30$  and $\SNR=100$. Abscissa  is normalised specific energy
$\eta$ and ordinate is specific angular momentum $h$.  Note that the maximum
residual error is well below   the signal's  amplitude and decreasing   with
increasing SNR.
\label{f:simu-f-var}}
\end{figure}

\subsection{Specificities of stellar disk inversion}

\subsubsection{Models of  azimuthal  velocity  distributions.}

Simulated  azimuthal  velocity  distributions  can  be  constructed  via the
prescription described in Pichon \& Lynden-Bell  \shortcite{PchLBd96}.   The
construction of Gaussian line profiles compatible with  a  given temperature
requires specifying the mean azimuthal velocity of  the  flow, $\left\langle
v_\phi\right\rangle$ on which the Gaussian should be  centered,  the surface
density, $\Sigma(R)$ and the  azimuthal  velocity  dispersion $\sigma_\phi$.
The line profile $F$ then reads
\begin{equation}
  F_\phi(R,v_\phi)= {\Sigma(R) \over \sqrt{2 \pi} \, \sigma_\phi } \,
  \exp\left( - {\left[ v_\phi- \left\langle v_\phi \right\rangle
  \right]^2 \over 2 \sigma_\phi^2} \right) \, . \label{e:formF2}
\end{equation}
\noindent 
Here the azimuthal velocity dispersion is related to the azimuthal pressure,
$ p_\phi$, by
\begin{equation}
  \sigma_\phi^2 = p_\phi/\Sigma - \left\langle v_\phi \right\rangle^2
  \, .
\end{equation}
\noindent The azimuthal pressure  $p_\phi$  follows  from  the  equation of
radial support,
\begin{equation}
  \left\langle v_\phi^2 \right\rangle - R {\partial \psi \over \partial R} =
  {\partial \left( R \Sigma \sigma_R^2 \right) \over \Sigma \, \partial R} \,,
\end{equation}
\noindent  and  the kinematical ``temperature'' of   the  disk with  a given
Toomre number $Q$ \cite{Toomre64}.
\begin{equation}
	Q=0.298 \, \sigma_R \kappa \,/ \, \Sigma \, . 
\end{equation}
  The  expression of the average azimuthal
velocity, $\langle v_\phi \rangle$, may be  taken to be  that which leads to
no asymmetric drift equation:
\begin{equation}
  \Sigma \left\langle v_\phi \right\rangle^2= p_\phi -
  p_R(\kappa \, R \, / \, 2 v_c)^2 \,,  \label{e:drift}
\end{equation}
\noindent  where  $\kappa$  is  the  epicyclic  frequency,  and  $v_{c}$ the
velocity of  circular  orbits.   Equations (\ref{e:formF2})--(\ref{e:drift})
provide a prescription for the  Gaussian  azimuthal  line  profile $F_\phi$.
These azimuthal  velocity  distributions  are  used  throughout  to generate
simulated data corresponding to iso-Q Kuzmin disk.

\subsubsection{The counter-rotating radial orbits}

As shown in \Fig{fig1},  the  azimuthal  distributions  of  our  models have
Gaussian tails corresponding to stars on  almost  radial  orbits  with small
negative azimuthal velocity.  These few stars play a  strong  dynamical role
in stabilising the disk, and as such should  not  be  overlooked  since they
significantly increase the azimuthal dispersion of inner orbits, effectively
holding the inner galaxy against its self-gravity.  Now  this  Gaussian tail
translates in the momentum, reduced energy space as a small group of counter
rotating orbits introducing a cusp in the number of stars  near  $h=0$ (this
cusp is only apparent  since  the  distribution  is  clearly  continuous and
differentiable across this line).  In practice the regularisation constraint
across $h=0$ is relaxed, treating in effect independently the two regions.

\subsection{Validation \& efficiency}

\subsubsection{Quality estimation}

Clearly the quality level for  the  reconstructed  distribution  will depend
upon  the  application  in  mind.   For  stability  analysis,  the  relevant
information involves for instance its gradients in action  space.   An acute
quality estimator  would  therefore  involve  such  gradients,  though their
computation requires some knowledge of the orbital  structure  of  the disk,
and  is  beyond  the  scope  of  this  paper.   Here  the  quality   of  the
reconstruction is estimated while computing  the  mean distribution-weighted
residual between the distribution sought and  the  model  recovered.   It is
defined by:
\begin{equation}
    \R{error}(\M{f})
    \!\!=\!\! \left\langle|f-f_\R{true}|\right\rangle 
    \!\!\simeq\!\! \frac{\sum_i f_{\R{true},i}\,|f_i - f_{\R{true},i}|}
           {\sum_i f_{\R{true},i}} \, , \\
\end{equation}
and measures the  restored  distribution  error  with  respect  to  the true
distribution   $\hat{f}_\R{true}(\eta,h)$    averaged    over    the   stars
(i.e.  weighted  by  the  distribution   $\hat{f}_\R{true}$).    a   set  of
simulations displaying this estimate for the quality  of  the reconstruction
was carried while varying respectively the outer sampled edge  of  the disk,
the signal to noise ratio, the sampling in the modeled distribution  and the
Q number of the underlying data set and is described bellow.

\subsubsection{Validation: zero noise level inversion}

An inversion without any noise is  first  carried  in  order  to  assess the
accuracy of our inversion routine.  This turned  out  to  be  more difficult
than doing the inversion with some knowledge of  the  noise  level  since in
this instance there is no simple assessment of a good value for the Lagrange
multiplier $\mu$.  All the ill  conditioning  arises  because  of  round off
errors alone.  The original distribution was  eventually  recovered  in this
manner  with  a  mean  distribution-weighted  residual,  $\R{error}(\M{f})$,
smaller than one part in ten to the four.   From  now  on,  the distribution
$\hat{f}(\eta,h)$ derived from this noise-free  inversion  is  taken  in our
simulations as the ``true'' underlying distribution.

\subsubsection{Choice of the penalising function}

Let us now  investigate  the  penalising  functions  corresponding  to three
methods of regularisation, namely MEM with uniform  prior  (as  advocated by
SB), MEM with smooth floating  prior  (given  by \Eqs{Rmem}{floating-prior})
and quadratic regularisation (\Eq{Rquad}).  The  corresponding  modelled and
recovered distribution functions are given  in  \Figs{fig1}  and \Fip{fig2}.
{}From these figures it appears  clearly  that  MEM  with  uniform  prior is
unsuitable in this context  (this  failure  is  expected  since  here  -- in
contrast  to  image  reconstruction  --  no  cutoff  frequency  forbits  the
roughness of the solution), while  MEM  with  floating  prior  or quadratric
penalizing function  --  which  both  enforces  smoothness,  provide similar
results and yield a satisfactory level of regularisation.  In particular, no
qualitative difference occur owing to the penalising  function  alone, which
is good indication that the inversion is carried adequately.  Note  that the
apparent cusp at $h=0$ is well accounted for by the inversion.

Regularisation by negentropy with a floating smooth prior  was  used  in the
following simulations.

\subsubsection{Efficiency: the influence of the noise level}

In the second part of the  simulations,  the  performances  of  the proposed
algorithm with respect to noise level is investigated.  The noise is assumed
to obey a Normal distribution with standard deviation given by:
\begin{equation}
\sigma_{i,j} = \frac{F_{i,j}}{\SNR} + 
\R{max}(\M{F})\, \sigma_\R{bg}\,.
\end{equation}
In other words,the intrinsic data  noise  has  a  constant  signal  to noise
ratio, $\SNR$, and that the detector adds a uniform readout noise.  Two sets
of  runs  corresponding  respectively  to  a  constant  signal  to  noise of
$\SNR=5$, $\SNR=30$, and $\SNR=100$ are presented in  \Fig{simu-f-avg} (mean
recovered    $\M{f}$),    \Fig{simu-f-sample}    (sample     $\M{f}$)    and
\Fig{simu-f-var} (standard deviation).   In  all  cases,  the  readout noise
level is $\sigma_\R{bg}=10^{-4}$.  The figures only display  the  inner part
of the distribution while the simulation carries the  inversion  for  $h$ in
the range $[-2,3[$ and all possible energies.

The main conclusion to be drawn from these figures is that the main features
-- both qualitatively and quantitatively given the  noise  level  --  of the
distribution are clearly recovered by this inversion  procedure.   Note that
near  the  peak  of  the  distribution  at  $h=0$,  $\eta=0$,  the recovered
distribution is nonetheless slightly rounder than  its  original counterpart
for the noisier ($\SNR=5$, panel 2) simulation.  This is a residual  bias of
the re-parametrisation: the sought distribution is  effectively undersampled
in that region and the regularisation truncates the residual  high frequency
in the signal while incorrectly assuming that it corresponds to  noise.   If
the sampling had been tighter in that region, say using regular  sampling in
$\exp(-\eta)$, the regularisation  would  not  have  truncated  the restored
distribution.  Alternatively, in order to retain algebraic  kernels, un-even
logarithmic sampling in the spline basis is an option.

This point illustrates the danger of non-parametric inversions which clearly
provide the best approach to model fitting  but  leave  open  some  level of
model-dependent tuning  and  consequently  potential  flaws  when  the wrong
assumptions are made on the nature of the sought solution for low  signal to
noise.  For instance, the above described procedure would  inherently ignore
any central cusp in the disk if  the  sampling  in  parameter  space  is too
sparse in that region, even if the signal  to  noise  level  is  adequate to
resolve  the  cusp.   Since  in  practice  a   systematic   oversampling  is
computationally onerous given the  dimensionality  of  the  problem, special
care  should  be  taken  while  deciding  what  an  adequate   sampling  and
parametrisation involves.

Finally, \Fig{fit-vs-snr} gives the evolution of  the  fit  error  signal to
noise ratio for various Toomre parameters $Q$.  This figure illustrates that
the method is independent of the model disk, dynamically cold or hot.

\subsubsection{Efficiency: sampling in the model}

\begin{figure}
\centerline{\psfig{file=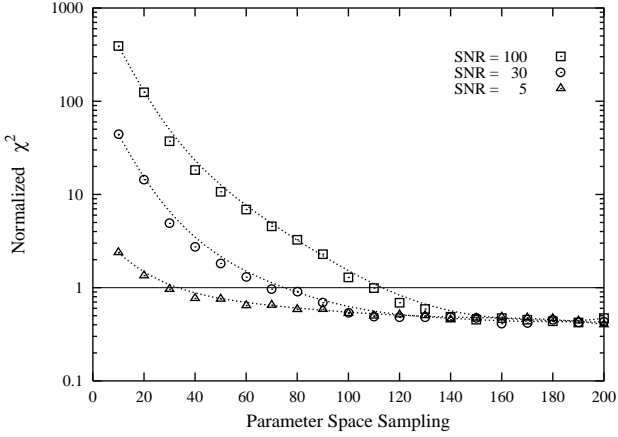,width=\columnwidth}}
\caption{Minimum value of the normalised likelihood term $\chi^2/N_\R{data}$
that can be  reached  as  the  number  of  basis  functions  varies  and for
different signal-to-noise  ratios.   In  abscissa  is  given  the  number of
samples along $\eta$ and $h$, which is the  square  root  of  the  number of
basis functions used.  The number of data measurements were $50\times50$ and
the maximum disk radius was $R_\R{max}=7$.   The  curves  are  only  here to
clarify the figure: the simulation results are plotted as symbols.}
\label{f:minimum-likelihood}
\end{figure}

The best sampling of the phase space of $f$ must be derived considering that
two opposite criteria should be balanced: (i) using too few  basis functions
would bias the solution, (ii) using more basis functions consumes  more {\sc
cpu} time.  A simple and intuitive way to check that  the  sampling  rate is
sufficient  is  to  insure  that  the  minimum  likelihood  reached  without
regularisation  is  much   smaller   than   the   target   likelihood,  i.e.
$\lim_{\mu\rightarrow0}L(\M{f})\ll  N_\R{e}$.   Unregularised  inversions of
noisy data with an increasing number of basis functions  and signal-to-noise
ratios was therefore performed.  In practice, since completely  omitting the
regularisation leads to a difficult minimisation problem because of  a large
number of local minima, regularisation was instead relaxed by using a target
likelihood   somewhat    lower    than    the    number    of   measurements
($N_\R{e}\simeq0.1\times N_\R{data}$).  The results of these simulations are
displayed    in    Fig.~\ref{f:minimum-likelihood}.     It    appears   that
$\sim150\times150$ basis functions  are  sufficient  to  avoid  the sampling
bias.  In all the other simulations, $150\times150$ or  $200\times200$ basis
functions are used.

\subsubsection{Efficiency: truncation in the  measurements}

\begin{figure}
\centerline{\psfig{file=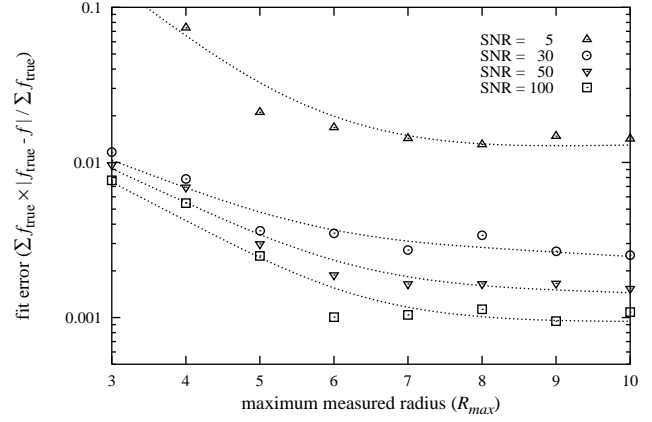,width=\columnwidth}}
\caption{Evolution of the fit error as the data set is  truncated  in radius
for different signal-to-noise ratios (in the simulations the disk outer edge
was assumed to be 10; the curves are only here to guide the eye: the results
of simulation are plotted as symbols).}
\label{f:best-fit-vs-rmax}
\end{figure}

The  inversion  algorithm  presented   here   makes   no   assumption  about
completeness of the input data set.  Therefore, the  recovered  solution $f$
can in principle be  used  to  predict  missing  values  in  $F_\phi$  -- in
contrast to direct inversion methods which  assume  that  $F_\phi$  is known
everywhere.  Real data will  always  be  truncated  at  some  maximum radius
$R\le{}R_\R{max}$. There may also be missing measurements  for  instance due
to dust clouds which hide some parts of the disk,  or  departure  from axial
symmetry  corresponding  to  spiral  structure.   In  order  to   check  how
extrapolation proceeds,  various  truncated  data  were  simulated  sets and
carried the inversion.  Figure~\ref{f:best-fit-vs-rmax} shows  the departure
of the recovered distributions from the true one as a function of  the outer
radius $R_\R{max}$ up to which data is  measured.   This  figure  also shows
that our inversion allows some extrapolation since  for  all signal-to-noise
ratios  considered,  the  error  reaches  it  minimum  value   as   soon  as
$R_\R{max}\ge7$ (i.e. 4 half mass radii, $R_{e}$, to be compared to the true
disk radius which was 10 in our simulations).   Note  that  interpolation is
likely to be more reliable than extrapolation; our  method  should therefore
be much less sensitive to data ``holes''.

\begin{figure}
\centerline{\psfig{file=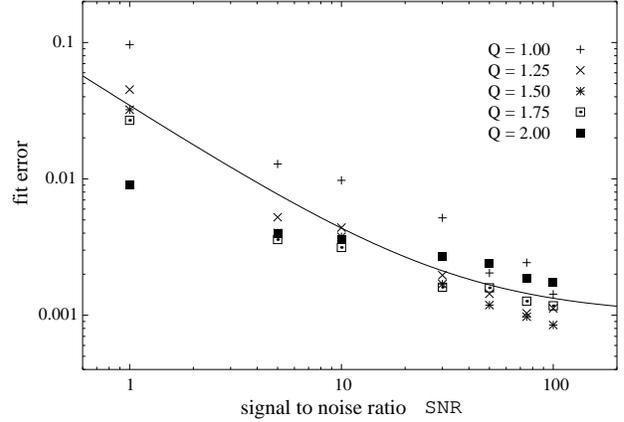,width=\columnwidth}}
\caption{Fit error versus SNR for various Toomre  parameters  $Q$.   The fit
error               approximatively               decreases              as:
$\R{error}\simeq1.0\times10^{-3}+3.4\times10^{-2}/\SNR$ (solid line curve).
\label{f:fit-vs-snr}}
\end{figure}

\section{Discussion \& Conclusion} \label{s:discussion}

This  paper  presented  a  series  of  practical  algorithms  to  obtain the
distribution   function   $f$   from   the   {\it   measured}  distributions
$F_\phi(R,v_\phi)$, compared those to existing algorithms  and  described in
details the best suited to carry efficient inversion of such ill-conditioned
problems.  It was argued that non-parametric modelisation is best  suited to
describe the underlying distribution functions when  no  particular physical
model is to be favoured.  For these inversions, regularisation is  a crucial
issue and its weight should be tuned ``on the fly'' according  to  the noise
level.


The mimimization algorithms  brought  forward  in section~\ref{section:algo}
are fairly general and could clearly be implemented to minimization problems
corresponding to other geometries such as that corresponding to the recovery
of distributions for  spheroid  or  elliptical  galaxies  explored  by other
authors (e.g. Merritt \& Tremblay \shortcite{DM87}).  More generally it could  be  applied to
any linear inversion problem where positivity  is  an  issue;  this includes
image reconstruction,  all  Abel  deprojection  arising  in  astronomy, etc.
Applying this algorithm to simulated  noisy  data,  it  was  found  that the
criteria of positivity and smoothness alone  are  sufficiently  selective to
regularize the inversion  problem  up  to  very  low  signal-to-noise ratios
($\SNR\sim5$) as soon as data is available up to 4  $R_{e}$.   The inversion
method described here is directly applicable to published measurements.

Here the inversion assumed that the HI rotation  curve  gives  access  to an
analytic (or spline) form  for  the  potential.   A  more  general procedure
should provide a simultaneous  recovery  of  the  potential,  though  such a
routine would be very CPU intensive since chanching  the  potential requires
to recompute the matrix $\M{a}$.  Nevertheless, it would  be straightforward
to extend the scope of this method  to  configurations  corresponding  to an
arbitrary slit angle as sketched in the Appendix~\ref{section:alpha},  or to
data produced  by  integral  field  spectroscopy  [such  as  TIGRE  or OASIS
\cite{BaconEtAl1995}] where the redundancy in azimuth would  lead  to higher
signal to noise  ratios  if  the  disk  is  still  assumed  to  be  flat and
axi-symmetric.

Once the distribution function has been  characterised,  it  is  possible to
study quantitatively  all  departures  from  the  flat  axisymmetric stellar
models.  Indeed, axisymmetric distribution functions are the building blocks
of all sophisticated stability analyses, and a good phase space  portrait of
the unperturbed configuration  is  clearly  needed  in  order  to  asses the
stability  of  a  given  equilibrium  state.   Numerical  N-body simulations
require sets of initial conditions which should reflect  the  nature  of the
equilibrium.  Linear stability analysis also rely on a detailed knowledge of
the underlying distribution\cite{PchCan97}.



\section*{Acknowledgements}
{\em We would like to thank R.~Cannon \& O.~Gerhard  for  useful discussions
and  D.~Munro  for  freely  distributing  his  Yorick  programming  language
(available at {\em\tt ftp://ftp-icf.llnl.gov:/pub/Yorick}) which we  used to
implement our algorithm  and  perform  simulations.   Funding  by  the Swiss
National  Fund  and  computer  resources  from   the   IAP   are  gratefully
acknowledged. Special thanks to J.~Magorrian for his help with Appendix~C.}

\def\BIBITEM#1#2#3{\bibitem[\protect\citename{#2, }#3]{#1}}

\def\aj {Astronomical Journal}

\appendix

\section{Bilinear interpolation}\label{section:linearInterpolation}

In  this  non-parametric  approach,  the  distribution  $\hat{f}(\eta,h)$ is
described by its projection onto a basis  of  functions.   If  we  choose a
basis for which  the  two  variables  $\eta$  and  $h$  are  separable then
Eq.~(\ref{e:f-model-2}) becomes:
\begin{equation}
	\hat{f}(\eta,h)=\sum_{k}\sum_{l}f_{k,l} \, u_k(\eta) \,
	v_l(h)\,,
	\label{e:f-model-3}
\end{equation}
\noindent where $u_k(\eta)$ and $v_l(h)$ are the new basis functions.  This
description of $\hat{f}(\eta,h)$ yields:
\begin{equation}
	\tilde{F}_\phi(R_i,{v_\phi}_j)
	= \tilde{F}_{i,j} = \sum_{k}\sum_{l} a_{i,j,k,l} \, f_{k,l}
	\,,\label{e:Fphi-model-3}
\end{equation}
\noindent where $a_{i,j,k,l}$ are coefficients which only depends on $R_i$,
${v_\phi}_j$ and $\psi_i$ ($\psi(R)$ in fact):
\begin{equation}
	a_{i,j,k,l} = v_l(R_i\,{v_\phi}_j) \sqrt{-2\epsmin_{i,j}}
		\int^1_{\etacrc_{i,j}}
		\frac{u_k(\eta)}{\sqrt{\eta-\etacrc_{i,j}}}\,\d{\eta}\,.
		\label{e:a-model-1}
\end{equation}
  Bilinear   interpolation  is implemented in the simulations
described in section~4  to  evaluate
$\hat{f}(\eta,h)$ everywhere.  In this case, the weights $f_{k,l}$  are the
values of the distribution  at  the  sampling  positions  $\{(\eta_k, h_l);
k=1,\dots,K;l=1,\dots,L\}$:
\begin{displaymath}
	f_{k,l}=\hat{f}(\eta_k,h_l) \,,
\end{displaymath}
and the basis functions are linear splines:
\begin{eqnarray*}
  u_k(\eta) &=& \left\{\begin{array}{ll}
  	1-\left\vert\frac{\eta-\eta_k}{\Delta\!\eta}\right\vert
  		& \mbox{if $\eta_{k-1}\le\eta\le\eta_{k+1}$,}\\
  	0
  		& \mbox{otherwise,}\\
  	\end{array}\right.
  \\
  v_l(h) &=& \left\{\begin{array}{ll}
  	1-\left\vert\frac{h-h_l}{\Delta\!h}\right\vert
  		& \mbox{if $h_{l-1}\le{}h\le{}h_{l+1}$,}\\
  	0
  		& \mbox{otherwise,}\\
  	\end{array}\right. 
\end{eqnarray*}
\noindent        with         $\eta_{k+n}=\eta_k+n\,\Delta\!\eta$        and
$h_{l+n}=h_l+n\,\Delta\!h$.  The bilinear interpolation is a particular case
of the general non-parametric description.  It yields a  very  sparse matrix
$\M{a}$ which  can  significantly  speed  up  matrix  multiplications.   The
coefficients $a_{i,j,k,l}$ can be computed  analytically,  though  since the
basis functions are defined piecewise,  the  integration  much  be performed
piecewise:
\begin{displaymath}
  \int_\etacrc^1\frac{u_k(\eta)}{\sqrt{\eta-\etacrc}}\,\d{\eta}
  	= \left\{\begin{array}{ll}
  		\alpha''_k		& \mbox{for $k=1$,}\\
  		\alpha'_k+\alpha''_k	& \mbox{for $k=2,\ldots,K-1$,}\\
  		\alpha'_k		& \mbox{for $k=K$,}\\
  	\end{array}\right.
\end{displaymath}
\noindent with:
\def\EtaLow{{\R{max}\!\{\eta_{k-1},\etacrc\!\}}}
\def\EtaHigh{{\R{min}\!\{\eta_{k},1\!\}}}
\begin{displaymath}
  \alpha'_k =
  	\int_\EtaLow^\EtaHigh\frac{u_k(\eta)}{\sqrt{\eta-\etacrc}}\,\d{\eta}
\, ,
\end{displaymath}
\begin{displaymath}
  \phantom{\alpha'_k} = \!\Biggl\{\!\!\!
	\begin{array}{ll}
	0 & \mbox{if $\eta_k\!\le\!\etacrc$,}\\
	\left[\frac{2\sqrt{\eta-\etacrc}}{3\Delta\!\eta}
	(\eta\!-\!3\eta_{k-1}\!+\!2\etacrc)
	\right]_{\eta=\EtaLow}^{\eta=\EtaHigh}
	& \mbox{otherwise.}\\
	\end{array}
\end{displaymath}
and
\def\EtaLow{{\R{max}\!\{\eta_{k},\etacrc\!\}}}
\def\EtaHigh{{\R{min}\!\{\eta_{k+1},1\!\}}}
\begin{displaymath}
  \alpha''_k =
  	\int_\EtaLow^\EtaHigh\frac{u_k(\eta)}{\sqrt{\eta-\etacrc}}\,\d{\eta}
  	\, ,
\end{displaymath}
\begin{displaymath}
  \phantom{\alpha''_k} = \!\Biggl\{\!\!\!
	\begin{array}{ll}
	0 & \mbox{if $\eta_{k+1}\!\le\!\etacrc$}\\
	\left[\frac{2\sqrt{\eta-\etacrc}}{3\Delta\!\eta}
	(3\eta_{k+1}\!-\!\eta\!-\!2\etacrc)
	\right]_{\eta=\EtaLow}^{\eta=\EtaHigh}
	& \mbox{otherwise.}\\
	\end{array}
\end{displaymath}

Another useful feature of  bilinear  interpolation  is  that  the positivity
constraint is straightforward to implement:
\begin{displaymath}
	\hat{f}(\eta,h)=\sum_{k,l}f_{k,l}u_k(\eta)v_l(h) \ge 0;
	\forall (\eta,h) \quad \Leftrightarrow \quad
	f_{k,l} \ge 0; \forall (k,l)\,.
\end{displaymath}
There is no such simple relation  for higher order splines.

\section{Specific minimisation methods for MEM}\label{s:mem}

Several non-linear  methods  have  been  derived  specifically  to  seek the
maximum entropy solution.  Let us review those briefly so as to compare them
with our method (section~\ref{s:pos1d}).   In  MEM,  assuming  that  (i) the
prior $\M{p}$ does not depend on the parameters and that (ii) the Hessian of
the likelihood term can be neglected, the  Hessian  of  $Q$  is  then purely
diagonal:
\begin{displaymath}
	\nabla\!\nabla\!Q_{k,l} \simeq \mu\,\nabla\!\nabla\!R_{k,l}
	= \frac{\mu\delta_{k,l}}{f_k}\,.
\end{displaymath}
The direction of minimisation is therefore:
\begin{equation}
	{\delta\!f_\R{MEM}}_k = -\frac{f_k}{\mu} \, \nabla\!Q_k\,.
	\label{e:delta-f-MEM}
\end{equation}
Skilling \& Bryan \shortcite{SkillingBryan84} discussed  further refinements
to speed  up  convergence.   Cornwell  \&  Evans \shortcite{CornwellEvans84}
approximated the Hessian $\nabla\!\nabla\!Q$ by neglecting  all non-diagonal
elements:
\begin{displaymath}
	\nabla\!\nabla\!Q_{k,l} \simeq \delta_{k,l}\frac{\mu}{f_k}
	+\delta_{k,l}\sum_{i,j}\frac{a_{i,j,k}^2}{\R{Var}(\tilde{F}_{i,j})}
	\, ,
\end{displaymath}
which yields:
\begin{equation}
	{\delta\!f_\R{CE}}_k = -\frac{f_k \, \nabla\!Q_k}{\mu +
	f_k\sum_{i,j}{a_{i,j,k}^2}/{\R{Var}(\tilde{F}_{i,j})}} \, . 
	\label{e:delta-f-CE}
\end{equation}
In fact, ${\delta\!f_\R{CE}}_k$ is equivalent to a steepest descent  step in
the Levenberg-Marquart method \cite{NumericalRecipes} which is the method of
predilection  to  fit  a  parametric   non-linear   model.    Extending  the
Richardson-Lucy method to the maximum  penalised  likelihood  r\'egime, Lucy
\shortcite{Lucy94} suggests:
\begin{equation}
	{\delta\!f_\R{Lucy}}_k = - f_k \left[\nabla\!Q_k
		- \frac{\sum_l f_l \nabla\!Q_l}{\sum_l f_l}\right] \, , 
	\label{e:delta-f-Lucy}
\end{equation}
which is almost the same as in classical MEM but for  the  term  $\sum_l f_l
\nabla\!Q_l/\sum_l f_l$ which accounts for the  constraint  that $\sum_kf_k$
should remain constant.  Note that it is sufficient to replace $\nabla\!Q_k$
in  Eq.~(\ref{e:step-single})  by  $\nabla\!Q_k  -\sum_l  q_l  \nabla\!Q_l /
\sum_l  q_l$  to  apply  a   further   constraint   of   normalisation.

With all these non-linear methods, it may be advantageous to  also  seek the
step      size      which       minimises      $Q(\M{f}+\lambda\delta\M{f})$
\cite{CornwellEvans84,Lucy94}.

\section{General model with arbitrary slit orientation}\label{section:alpha}

\begin{figure}
\centerline{\psfig{file=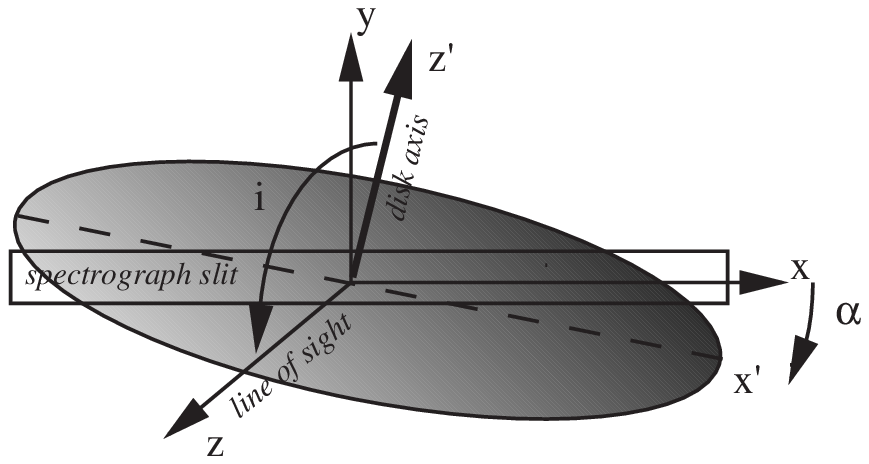,width=0.8\columnwidth}}
\caption{Angular notations\label{figaxe}}
\end{figure}

Long slit   spectroscopic observations  of   a galactic   disk provide  the
distribution:
\begin{equation}
  F_\alpha(R,v_\paral)= \int f(\eps,h)\,\d{v_\perp} \,,
  \label{e:Falpha}
\end{equation}
\noindent  where $v_\paral$  and   $v_\perp$ are   the star  velocities  (at
intrinsic radius,$r$ and projected radius  $R$)  along and perpendicular  to
the line of sight respectively, that are related to the radial and azimuthal
velocities:
\begin{displaymath}
  v_R=c_1\,v_\paral  +  c_2\,v_\perp  \,   , \quad   v_\phi=c_3\,v_\paral  +
    c_4\,v_\perp \quad \mbox{and}\quad R = c_5\ r \, .
\end{displaymath}
\noindent Here the $c_k$ depend on the angle $\alpha$ between the  slit and
the major axis of the disk as measured in the plane of the sky and  on
the inclination $i$ of the disk axis with respect to the line of  sight (see
Fig.~\ref{figaxe}).  The case where the slit is parallel to  the  major axis
of the disk, i.e.\ $\alpha=0$, as been examined  in  the  main  text.   When
$\alpha\not=0$, the specific angular momentum $h=r\,v_\phi$ can  be  used as
variable of integration:
\begin{equation}
  F_\alpha(R,v_\paral)= \frac{c_5}{R\,c_4}
        \int_{-\Hmax}^{\Hmax} f(\eps,h) \,\d{h}.   \label{e:Falpha-2}
\end{equation}
\noindent where the specific energy and the integration bounds are:
\begin{eqnarray*}
  \eps &=& \frac{1}{2} v_\paral^2 
	  +\frac{1}{2}\left(\frac{h \, c_5}{R\,c_4}-\frac{c_3}{c_4}\,v_\paral\right)^2
	  -\psi\left(\frac{R}{c_5}\right) \label{e:Falpha-en} \,,	\\
  \Hmax &=& \frac{R}{c_5} \left( \frac{c_3}{c_4}\,v_\paral
  	+  \sqrt{2\psi\left(\frac{R}{c_5}\right)-v_\paral^2} \right)  \label{e:Falpha-H2}\,.
\end{eqnarray*}
In practice, straightforward trigonometry yields
\begin{eqnarray*}
 c_1 &=& \sin(\beta) /\sin(i)\, ,\,\, 
 c_2 = \cos(\beta)\, ,\,\, \\
 c_3 &=& \cos(\beta) /\sin(i)\, ,\,\, 
 c_4 = \sin(\beta)\, ,\,\, 
\end{eqnarray*}
\[ c_5 = \sqrt{\cos^2(\beta)+\sin^2(\beta) \sin^2(i)}\,, \]
where $\beta$  is the angle of the slit as measured in the plane of the disk
which obeys
\begin{eqnarray*}
  \tan(\beta) &=& \tan(\alpha)/\sin(i) \, .
\end{eqnarray*}

\label{lastpage}
\end{document}